\documentclass[10pt,twocolumn,twoside,letterpaper]{IEEEtran}
\usepackage{enumerate}
\usepackage{subfigure}
\usepackage{url}
\usepackage{array}
\usepackage{amsmath}
\usepackage{amssymb}
\usepackage{amsfonts}
\usepackage{soul}
\usepackage{float}
\usepackage{multirow}
\usepackage{multicol}
\usepackage{extarrows}
\usepackage{mathrsfs}
\usepackage{bm}
\usepackage{booktabs}
\usepackage[sort]{cite}
\usepackage[table]{xcolor}
\usepackage{bigstrut}
\usepackage{arydshln}
\usepackage{ulem}
\usepackage{diagbox}
\usepackage[pdftex]{graphicx}

\graphicspath{{./Images/}}
\usepackage{epstopdf}
\DeclareGraphicsExtensions{.eps}

\newcommand{\red}[1]{\textcolor{red}{{#1}}}

\newcommand{\BL}[1]{\textcolor{blue}{{#1}}}

\usepackage{caption}
\begin{document}

\title{{Improving Cost Learning for JPEG Steganography by Exploiting JPEG Domain Knowledge}}

\author{
Weixuan Tang,
        Bin Li*,
        Mauro Barni,
        \\Jin Li,
       and Jiwu Huang,

\thanks{W. Tang and J. Li are with Institute of Artificial Intelligence and Blockchain, Guangzhou University, Guangdong 510006, China (email: tweix@gzhu.edu.cn; jinli71@gmail.com).}
\thanks{B. Li, and J. Huang are with Guangdong Key Laboratory of Intelligent Information Processing and Shenzhen Key Laboratory of Media Security, Shenzhen University, Shenzhen 518060, China (email: libin@szu.edu.cn; jwhuang@szu.edu.cn).}
\thanks{M. Barni is with Department of Information Engineering and Mathematics, University of Siena, Siena 53100, Italy (email: barni@dii.unisi.it).}
\thanks{*B. Li is the correspondence author.}
}
\maketitle

\begin{abstract}
Although significant progress in automatic learning of steganographic cost has been achieved recently, existing methods designed for spatial images are not well applicable to JPEG images which are more common media in daily life.
The difficulties of migration mostly lie in the unique and complicated JPEG characteristics caused by $8\times8$ DCT mode structure.
To address the issue, in this paper
{we extend an existing automatic cost learning scheme to JPEG, where the proposed scheme called JEC-RL (JPEG Embedding Cost with Reinforcement Learning) is explicitly designed to tailor the JPEG DCT structure.}
It works with the embedding action sampling mechanism under reinforcement learning, where a policy network learns the optimal embedding policies via maximizing the rewards provided by an environment network.
The policy network is constructed following a {domain-transition} design paradigm,
where three modules including {pixel-level texture complexity evaluation, DCT feature extraction, and mode-wise rearrangement,}
are proposed.
These modules operate in serial,
gradually extracting useful features from a decompressed JPEG image and converting them into embedding policies for DCT elements,
while considering JPEG characteristics including inter-block and intra-block correlations simultaneously.
The environment network is designed in a gradient-oriented way to provide stable reward values
by using a wide architecture equipped with a fixed preprocessing layer with $8\times8$ DCT basis filters.
Extensive experiments and ablation studies demonstrate that the proposed
method can achieve good security performance for JPEG images against both advanced feature based and modern CNN based steganalyzers.

\end{abstract}

\section{Introduction}
\label{sec:intro}

Image steganography and image steganalysis 
are a pair of antagonistic techniques, wherein the former conceals secret messages within cover images, and the latter looks for embedding artifacts to reveal the presence of secret messages within stego images.
Since JPEG is the most common image format and widely used in daily life, steganography and steganalysis for JPEG images are of both academic and practical value.

Modern JPEG steganographic methods are designed according to a distortion minimization framework \cite{distortion}. The distortion can be simply represented in an additive form, which is calculated as the sum of the embedding costs of modified DCT (Discrete Cosine Transform) coefficients. As a result, defining embedding cost is very important for steganography. In the past decade, some additive cost functions \cite{UNIWARD, UED, UERD, GUED, MS, JMIPOD} have been proposed either heuristically or based on a statistical model.
\if
In MOD (Model Optimized Distortion) \cite{MOD}, the embedding costs are defined associating with CC-PEV (Cartesian Calibration-PEV) steganalytic features \cite{CCPEV}. However, the costs optimized according to a specific model may be over-fitted and may not perform well when a steganalyzer utilizes a more complete feature set. In J-UNIWARD (JPEG-Universal Wavelet Relative Distortion) \cite{UNIWARD}, image residuals are obtained by filtering the decompressed image with a bank of wavelet filters, and then integrated into embedding costs. Although not only non-zero AC (Alternating Current) coefficients but also DC (Direct Current) coefficients and zero ACs are modified, J-UNIWARD achieves better security performance compared to previous methods. To avoid the high computational complexity in wavelet domain, UED (Uniform Embedding Distortion) \cite{UED}computes embedding costs in the DCT domain and makes embedding changes uniformly spread in DCT coefficients of different magnitudes. Moreover, it is important to design embedding costs associated with the correlation within the same DCT block and that among different DCT blocks, as these two characteristics are commonly captured by JPEG steganalysis. Following this principle, UERD (Uniform Embedding Distortion Revisited) \cite{UERD} and GUED (Generalized Uniform Embedding Distortion) \cite{GUED} define embedding costs by considering the impacts of different DCT frequency-modes and different DCT blocks.
To explore texture information more delicately, microscope technique is utilized in J-MSUNIWARD (Microscope JPEG-Universal Wavelet Relative Distortion) and MSUERD\_SPA (Microscope Uniform Embedding Distortion Revisited Filtering in Spatial Domain) to refine the existing costs by means of highlighting the details of the image with high-pass filters \cite{MS}.
\fi
Most existing well-performed embedding costs exploit {JPEG DCT characteristics, including the texture level of DCT blocks, the correlation among DCT coefficients, and the different impacts of DCT frequency-modes.} These additive embedding costs can be adjusted into non-additive embedding costs \cite{BBC++} or side-informed embedding costs \cite{sideinfor}.
On the other side, many efficient JPEG steganalytic methods are designed in a supervised machine learning fashion where high-dimensional hand-crafted image statistical features \cite{JRM, GFR,PHARM,diversity,selection} were constructed.
Their performance can be further improved by incorporating selection-channel information \cite{SCA,JPEGphaseaware}.

In recent years,
with the rapid development of deep learning techniques, steganalytic methods based on CNNs (Convolutional Neural Networks), including PNet \cite{PNet}, J-XuNet \cite{XuJPEG}, Zeng-CNN \cite{zeng}, and SRNet \cite{SRNet},  have achieved good performance.
To overcome the great challenges presented by deep learning aided steganalysis,
it is also tempting to take the advantages of deep learning for steganography so as to obtain better steganographic security.
In fact, deep machine learning can be applied in both discriminative and generative tasks.
There are already some good deep learning based data hiding methods available \cite{baluja2017hiding},
however, they can neither resist advanced steganalytic or forensic detection, nor achieve
error-free message decoding.
A compromised feasible  solution is to learn better embedding costs under the distortion minimization framework.
Compared to traditional cost functions predefined by heuristics or a fixed model,
learning costs from scratch by deep learning is appealing
for that it can automatically learn intrinsic cover generative representation from big data and be adjusted dynamically on demand.




There are two types of state-of-the-art techniques for automatic embedding cost learning.
One is generative adversarial network (GAN) based and the other is
reinforcement learning (RL) based.
In the first type,
the earliest work is ASDL-GAN (Automatic Steganographic Distortion Learning framework with Generative Adversarial Network) \cite{ASDLGAN},
in which
a generator is designed to learn embedding change probabilities
so as to resist a discriminator which aims to distinguish between cover and simulated stego images.
A neural network based embedding simulator is used to simulate message embedding.
UT-GAN (U-Net and Double-Tanh Framework using GAN) \cite{UTGAN} has improved ASDL-GAN by utilizing more advanced network architectures and a double-Tanh activation function based simulator.
In the second type,
SPAR-RL (Steganographic Pixel-wise Actions and Rewards with Reinforcement Learning) \cite{SPARRL}
has been proposed
by using a sampling based simulator to sample embedding actions
so as to overcome the ``trade-off drawback'' (namely, the drawback of trading-off accurate simulated modifications for attenuated gradients) caused by neural network based (or activation function based) embedding simulators in ASDL-GAN/UT-GAN.
In SPAR-RL, a policy network aims to generate embedding policies, i.e., embedding change probabilities, by maximizing the rewards assigned from an environment (or called critic) for the sampled modification actions.
There is yet another related technique employing deep learning for steganography called adversarial embedding (ADV-EMB) \cite{ADVEMB}.
It imitates the effect of adversarial examples (AE) by utilizing the gradients of a CNN steganalyzer
 for adjusting costs to evade detection.
{A min-max strategy \cite{minmax} can further be applied to construct a set of trained steganalyzers so that a min-max game equilibrium can be gradually obtained to optimize the performance of the adjusted costs.}
However, such technique can only adjust the off-the-shelf embedding costs while it is incapable of
generating costs from scratch.
In fact, it can be used as an adjoint technique as non-additive cost functions \cite{BBC++} to further boost the steganographic performance, which has been demonstrated in \cite{SPARRL} to improve the costs generated by SPAR-RL.
The idea of learning embedding costs in single image steganography can be further extended to batch steganography \cite{batch}.

Although some effective automatic cost learning methods have been proposed so far,
the methods are only applicable to spatial images for that the costs are explicitly defined on pixels rather than DCT coefficients.
They cannot be well applied on JPEG images for that the very embedding units reside in DCT domain.
{JS-GAN (JPEG Steganography using GAN) \cite{JSGAN} is the earliest method attempted to automatically learn JPEG embedding costs.
But its network architecture does not fully consider the peculiarity of JPEG images and its activation function based embedding simulator has the ``trade-off drawback'' mentioned above. Therefore, its performance is still inferior to conventional methods such as the baseline J-UNIWARD (JPEG-Universal Wavelet Relative Distortion) \cite{UNIWARD}.

Note that most existing JPEG steganographic methods \cite{UED, UERD, GUED, MS} except the one from the UNIWARD family \cite{UNIWARD} are designed to tailor JPEG characteristics.
The special DCT $8\times8$ mode structure is the major source of difficulties that hinder
a successful migration from a spatial pixel-oriented method to a JPEG DCT-oriented one.
This fact may also hold true for steganography based on deep learning.
For instance, both inter-block correlations and intra-block correlations exist among DCT coefficients, which are not easily captured by conventional convolution operations presented in previous automatic cost learning methods \cite{ASDLGAN,UTGAN,SPARRL,JSGAN}.

To address the above problems, in this paper we extend the RL based automatic embedding cost learning scheme with a sampling based simulator to DCT domain and propose a method called JEC-RL (JPEG Embedding Cost with Reinforcement Learning).
Our method explicitly exploits JPEG domain knowledge utilized in conventional steganography and steganalysis.
Such domain knowledge includes the embedding cost measurement from two aspects in conventional steganography \cite{UNIWARD,UERD,GUED,MS},
i.e., the texture complexity of a DCT block and the
position of its DCT frequency-mode,
and also includes the important aspect in modern JPEG steganalysis \cite{JRM, GFR,PHARM,SCA,PNet,XuJPEG,zeng,SRNet,selection,JPEGphaseaware,diversity}, i.e., the intra-block and inter-block correlations of DCT elements.
In this paper, we propose a three-module composed policy network to consider these factors.
These modules operate in serial,
gradually extracting useful features from a decompressed JPEG image and converting them into embedding policies for DCT elements.
Moreover, since all DCT coefficients are possible to be modified during data embedding,
many deep CNN steganalyzers equipped with high-pass filters \cite{Xu, Yedroudj} may not be suitable to be adopted as an environment network for assigning rewards.
We investigate two properties that an effective neural network based environment may follow.
First, $8\times8$ DCT basis filters are better than high pass filters to be used in a preprocessing layer {to provide sufficient frequency resolution}.
Second, a ``wide" network is better than a ``deep" network for efficient gradient propagation and thus leading to better reward assignment.

The contributions and technical highlights of this work are summarized as follows.
\begin{itemize}
  \item A practical automatic cost learning method named JEC-RL has been proposed based on JPEG domain knowledge. Experiments show that JEC-RL can learn effective costs that
      outperform existing additive cost functions against feature-based and CNN-based steganalyzers. This is the first work that an automatic cost learning method starting from scratch can achieve outstanding performance for JPEG images.
      {Furthermore, the data-driven learning architecture can make better use of the texture calculation process in conventional JPEG steganographic methods.}

  \item
      A policy network for generating embedding policies has been constructed with three modules following a domain-transition design paradigm.
  In this paradigm,
  texture complexity is firstly evaluated in spatial domain by a \textit{pixel-level texture complexity evaluation module}.
      Then, by means of a \textit{DCT feature extraction module}, DCT features with inter-block and intra-block correlations can be learned in an end-to-end learning fashion.
      Finally, through a \textit{mode-wise rearrangement module}, DCT features are rearranged into $8\times8$ block structure as embedding policies.
  \item A gradient-oriented environment network has been adopted to facilitate reward assignment.
      Extensive ablation studies show that utilizing a preprocessing layer equipped with a bank of $8\times8$ DCT basis filters can {sense the modifications in all DCT modes so as to provide sufficient frequency resolution}.
      Besides, extensive ablation studies show that increasing the network capacity by widening the structure can enable stable gradient propagation so as to provide discriminative rewards.
\end{itemize}

This paper is organized as follows. In Section \ref{sec:fundamentals}, we give fundamental knowledge related to the proposed steganographic method, including the existing cost functions for JPEG images and the cost learning methods with deep learning techniques.
In Section \ref{sec:SPARRLv1v2}, we analyze and experimentally verify the ineffectiveness of existing automatic cost learning methods for JPEG images.
In Section \ref{sec:method}, we introduce the proposed JEC-RL by giving the details of its design paradigm and network architecture. In Section \ref{experiment}, we present extensive experimental results to demonstrate the performance.
In Section \ref{sec:conclusion}, we draw conclusions.

\section{Fundamentals And Background}
\label{sec:fundamentals}

\subsection{Notations}
\label{Sec: notation}

In the remaining of this paper, capital bold letters and the corresponding lowercase letters respectively represent the matrices and the elements within matrices. Specifically, the spatial grayscale cover and stego images are denoted as $\mathbf X = (x_{i,j})^{H \times W }$ and $\mathbf Y = (y_{i,j})^{H \times W }$, respectively, where $H$ and $W$ are the height and width of the image.
Without loss of generality, assume $H$ and $W$ are the multiples of 8.
The JPEG grayscale cover and stego images are respectively denoted as $\mathbf X^{J} = (x_{a,b}^{k,l})^{H \times W }$ and $\mathbf Y^{J} = (y_{a,b}^{k,l})^{H \times W }$, where $1 \leq a \leq  H/8 $, $1 \leq b \leq W/8 $, $1 \leq k,l \leq 8$. Note that
$x_{a, b}^{k, l}$ (or $y_{a, b}^{k, l}$) is the
$\big(8\times(a-1)+k,8\times(b-1)+l\big)$-th element in $\mathbf X^{J}$ (or $\mathbf Y^{J}$), which corresponds to the DCT coefficient in the $(a,b)$-th DCT block and the $(k,l)$-th DCT frequency-mode.
Without loss of generality, we assume $H=W=256$.

\subsection{Basics of Embedding Costs}
\label{Sec: JPEGrho}
According to distortion minimization framework \cite{distortion}, the message embedding process can be formulated as an optimization problem with a payload constraint given as follows
\begin{equation}\label{equ:problem}
\min \limits_{\mathbf{Y}^{J} } D(\mathbf{X}^{J}, \mathbf{Y}^{J} ),  \quad
\text{s.t. }  \psi(\mathbf{Y}^{J} ) = C,
\end{equation}
where $D(\mathbf{X}^{J}, \mathbf{Y}^{J})$ is a function measuring the overall distortion
caused by modifying $\mathbf{X}^{J}$ to $\mathbf{Y}^{J}$, $\psi(\mathbf{Y}^{J})$ is the payload conveyed by $\mathbf{Y}^{J}$, and $C$ is the target payload.
A distortion function defined in an additive form can be expressed as
\begin{equation}\label{equ:add_distortion_all}
D({\mathbf{X}^{J}, \mathbf{Y}^{J}}) = \sum_{a=1}^{H/8} \sum_{b=1}^{W/8} \sum_{k=1}^{8} \sum_{l=1}^{8} \rho_{a,b}^{k,l}(m_{a,b}^{k,l})\mathcal{I}(m_{a,b}^{k,l}\neq0),
\end{equation}
where
$\rho_{a,b}^{k,l}(m_{a,b}^{k,l})$ is the additive embedding cost of modifying $x_{a,b}^{k,l}$ into $y_{a,b}^{k,l}=x_{a,b}^{k,l}+m_{a,b}^{k,l}$, and $\mathcal{I}(z)$ is the indicator function as
\begin{equation}\label{equ:delta}
    \mathcal{I}(z)=
    \begin{cases}
        1, \quad \text{if } z \text{ is true},\\
        0, \quad \text{if } z \text{ is false}.
    \end{cases}
\end{equation}
{In general, the more risk the modification may introduce, the larger the cost value is.}
Steganographic codes such as STC (Syndrome-Trellis Codes) \cite{STCs} can be applied in practice.
For simulation purpose,
an optimal embedding simulator \cite{simulator} can be used to compute the embedding change probabilities under a given payload as
\begin{equation}\label{eq:p+}
\begin{aligned}
p_{a,b}^{k,l}(m) = \frac{e^{-\lambda\rho_{a,b}^{k,l}(m)}}{ \sum_{\tilde{m} \in \mathcal{M}} e^{-\lambda\rho_{a,b}^{k,l}(\tilde{m})}}, \quad m \in \mathcal{M},
\end{aligned}
\end{equation}
and $\lambda$ is a parameter determined by the payload constraint which can be computed as
\begin{equation}
     -\sum_{a=1}^{H/8}\sum_{b=1}^{W/8}\sum_{k=1}^{8}\sum_{l=1}^{8}
    \sum_{m \in \mathcal{M}}
    p_{a,b}^{k,l}(m) \text{log}_2 p_{a,b}^{k,l}(m) =C.
\label{equ:capacity}
\end{equation}
In this paper, we focus on the case of ternary embedding, wherein
$\mathcal{M}=\{+1,0,-1\}$, and suppose $\rho_{a,b}^{k,l}(+1) = \rho_{a,b}^{k,l}(-1) = \rho_{a,b}^{k,l}$, and $\rho_{a,b}^{k,l}(0) = 0$.

\subsection{Embedding Costs in Conventional JPEG Steganography}
\label{Sec: JPEGrho2}
Generally speaking, in many existing JPEG steganographic methods, the embedding cost is measured from two aspects \cite{UERD},
including the texture complexity of a DCT block and the position of its DCT frequency-mode:
\begin{equation}\label{eq:JPEGrho}
\begin{aligned}
\rho_{a,b}^{k,l} = \rho_{a,b}^{<\text{block}>}.\rho_{k,l}^{<\text{mode}>}
\end{aligned}
\end{equation}
wherein $\rho_{a,b}^{<\text{block}>}$ is the block-level suitability of the $(a,b)$-th DCT block and $\rho_{k,l}^{<\text{mode}>}$ is the mode-level suitability of the $(k,l)$-th DCT frequency-mode.
Please note that $\rho_{a,b}^{<\text{block}>}$ can be set as the reciprocal of the block-level texture complexity and thus is closely related to the image content, while $\rho_{k,l}^{<\text{mode}>}$ relies on the position of the DCT frequency-mode $(k,l)$ and is independent of the image content.
In conventional methods \cite{UNIWARD,UERD,GUED,MS}, $\rho_{a,b}^{<\text{block}>}$ and $\rho_{k,l}^{<\text{mode}>}$ are computed independently.

Take UERD \cite{UERD} for example. The block-level suitability $\rho_{a,b}^{<\text{block}>}$ is defined as the reciprocal of the weighted sum of the block energy of the $(a,b)$-th DCT block and its neighboring blocks, given as
\begin{equation}\label{eq:UERDrho}
\begin{aligned}
\rho_{a,b}^{<\text{block}>} =
        \frac{1}{E_{a,b}+0.25\cdot \sum_{\hat{E}\in \mathbb{\hat{E}}_{a,b}}\hat{E}},
\end{aligned}
\end{equation}
where
$ \mathbb{\hat{E}}_{a,b}$ is the set of the block energy for the blocks located in the eight-neighborhood of the $(a,b)$-th DCT block,
and
the block energy $E_{a,b}$ of the $(a,b)$-th DCT block is defined as
\begin{equation}
\begin{aligned}
\label{eq:E}
E_{a,b} = \sum_{k=1}^{8}\sum_{l=1}^{8}|x_{a,b}^{k,l}| \cdot s_{k,l},
\end{aligned}
\end{equation}
where $s_{k,l}$ is the quantization step of the $(k,l)$-th DCT frequency-mode.
The mode-level suitability
$\rho_{k,l}^{<\text{mode}>}$ is defined according to the quantization steps $s_{k,l}$ as
\begin{equation}\label{eq:moderho}
\begin{aligned}
\rho_{k,l}^{<\text{mode}>} = \begin{cases}
        0.5\cdot(s_{k+1,l}+s_{k,l+1}), \quad  \text{if} \quad (k,l) = (1,1),\\
        s_{k,l},  \quad \text{otherwise}.
    \end{cases}
\end{aligned}
\end{equation}
The overall embedding costs $\rho_{a,b}^{k,l}$ is obtained according to \eqref{eq:JPEGrho}.

\begin{figure}[tb!]
\centering
\includegraphics[width=0.48\textwidth]{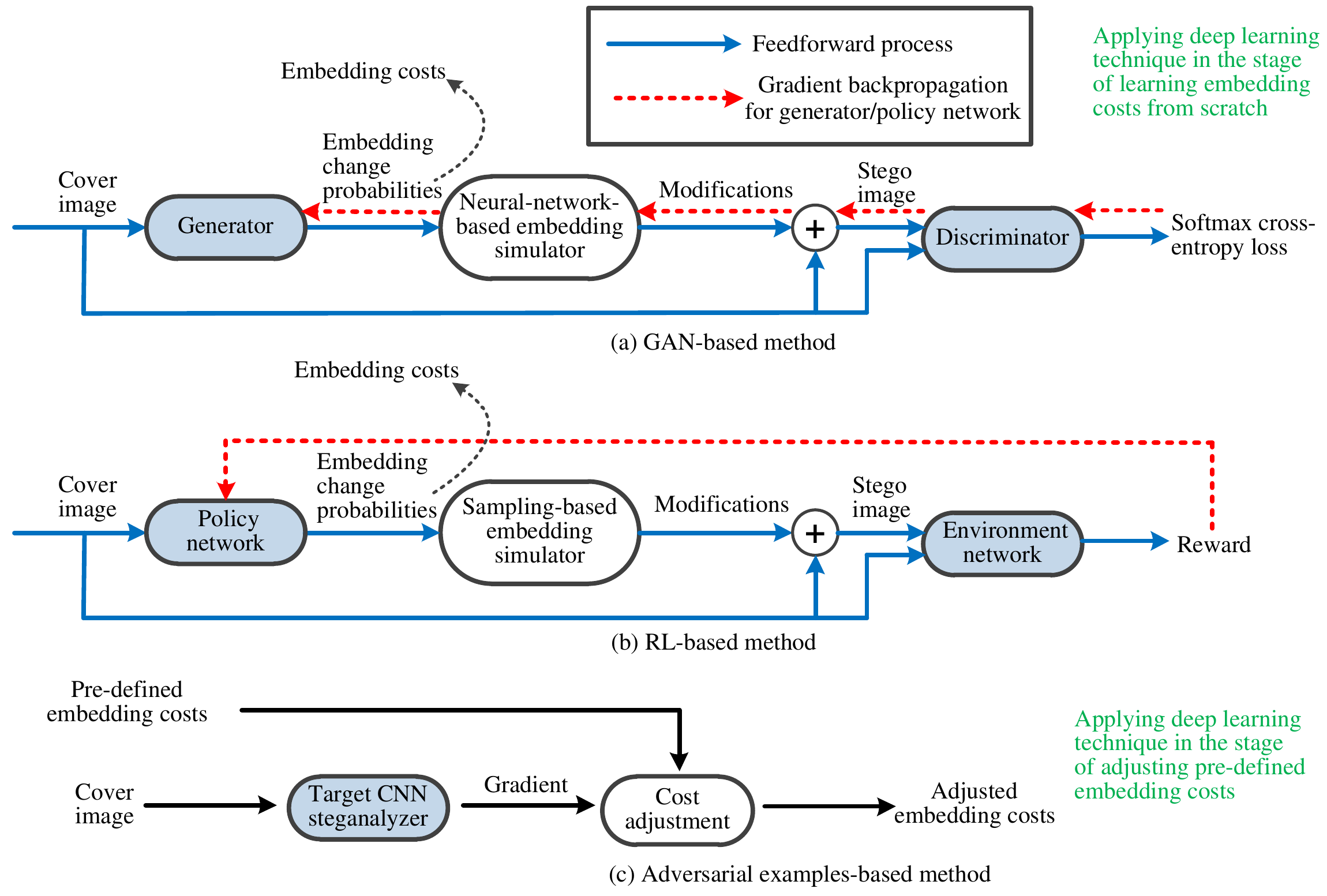}
\caption{
Three deep learning techniques for embedding costs.
}\label{fig:learn costs}
\vspace{-0.8cm}
\end{figure}

\subsection{Using Deep Learning for Embedding Costs}
\label{Sec: JPEGrho2}
There exist three kinds of deep learning techniques developed for embedding costs in steganography, including generative adversarial network (GAN) \cite{ASDLGAN, UTGAN, JSGAN}, reinforcement learning (RL) \cite{SPARRL}, and adversarial examples (AE) \cite{ADVEMB,zhang2018adversarial,minmax}.
Fig. \ref{fig:learn costs} shows the diagrams of their respective working flow.
They may share some similarities in network structure,
and their convergence status all reach a kind of equilibrium\footnote{This statement holds when an environment network is used in \cite{SPARRL} and a min-max strategy is used in \cite{minmax}. However, when a fix environment is used or the min-max strategy is disabled, the steganalysis side is considered as static.},
but they are different in their working mechanism.
To better understand their differences and why RL is adopted in our proposed method,
we compare these techniques side-by-side in details from the following aspects.
\begin{itemize}
  \item \textit{Scope of application:} Both GAN-based and RL-based method can be applied to automatically learn embedding costs from scratch, while AE-based methods are used to improve off-the-shelf embedding costs.
      In other words, GAN-based and RL-based methods can work independently while
      AE-based methods work as a kind of adjoint post-processing technique.

%

  \item \textit{Construction:} In GAN-based methods, a generator and a discriminator work together in the training stage to compete with each other, and only the generator is needed in the deployment stage to generate embedding costs. Both the generator and the discriminator are implemented by learnable neural networks.
      In AE-based methods, there should be a well-trained differentiable steganalyzer (or a set of steganalyzers in \cite{minmax}) for adjusting predefined embedding costs.
      In RL-based method, there exists a policy network and an environment (or called critic).
      In principle, the actions are sampled according to the embedding policies outputted by the policy network, and
      the environment can be any flexible module to yield rewards for the sampled actions.
      In practical implementation of \cite{SPARRL}, both policy network and the environment are
      realized by learnable neural networks.

  \item \textit{The usage of gradient:} In GAN-based methods,
  the gradient with respect to the network parameters
  is used to update the parameters of both the discriminator and the generator.
  In AE-based methods, the gradient with respect to the image elements
  is used to asymmetrically adjust the embedding costs.
  In RL-based method, the gradient with respect to environment network
  is used to update the parameters of the environment network,
  while such gradient is also incorporated into the reward function to guide the update of the parameters of the policy network.
  Note that if a non-gradient based reward function is designed, the gradient may not be necessarily involved in the update process of the policy network.

  \item \textit{Generating simulated stegos during the training stage:}
  In GAN-based methods,
  the simulated stegos are generated by a neural network based or activation function based simulator, where the simulated modifications are inaccurate floating values.
  In the RL-based method,
  the simulated stegos are sampled according to the embedding policies (probabilities),
  where the simulated modifications are accurate integer values.
  In the AE-based methods, genuine stegos are generated with an optimal embedding simulator
  or a practical stego code as mentioned in Section \ref{Sec: JPEGrho} according to the adjusted costs.
  In \cite{minmax}, several differentiable steganalyzers must be successively trained by a corresponding updated stego image set. Note that in the AE-based methods, since the input costs are predefined by a cost function or obtained by GAN-based or RL-based methods,
  the updating processes for adjusted costs and for the differentiable steganalyzers
  have much less number of training rounds (typically 8 or 9 in \cite{minmax}) compared to
  the GAN-based or RL-based methods.
    Both the discriminator in GAN and the environment network in RL are updated in a mini-batch fashion by
  using only several pairs of cover and simulated stego images,
  while the steganalyzers in AE must be updated by all training genuine stego images.

\end{itemize}

\if
The first work that can automatically learn embedding costs from scratch is ASDL-GAN (Automatic Steganographic Distortion Learning framework with Generative Adversarial Network) \cite{ASDLGAN}, which is designed with a GAN structure.
UT-GAN (U-Net and Double-Tanh Framework using GAN) \cite{UTGAN} has improved ASDL-GAN by utilizing a more sophisticatedly designed network architecture. With the help of the non-differentiable computation mechanism in reinforcement learning (RL), SPAR-RL (Steganographic Pixel-wise Actions and Rewards with Reinforcement Learning) \cite{SPARRL} is a new automatic cost learning framework \BL{with an embedding action sampling
mechanism}
that can overcome the dilemma of gradient vanishing and modification deviation in
neural-network-based embedding simulators in ASDL-GAN/UT-GAN. It utilizes
a policy network to learn optimal embedding policies associated with embedding costs via maximizing the reward assigned by an environment network.

\red{In recent years, deep learning techniques have been applied to learn embedding costs in steganography. As shown in Fig. \ref{fig:learn costs},  these methods can be divided into two stages, including learning embedding costs from scratch by generative adversarial network (GAN) or reinforcement learning (RL), and adjusting pre-defined embedding costs by adversarial examples technique.
As for the first stage, the earliest work is ASDL-GAN (Automatic Steganographic Distortion Learning framework with Generative Adversarial Network) \cite{ASDLGAN}.
In ASDL-GAN, the generator tires to learn embedding change probabilities for input cover image, which are then fed into a neural-network-based embedding simulator to obtain modifications. The modification map is added to the cover image as stego image. The discriminator aims to distinguish between cover and stego images.
The goal of the generator is to maximize the softmax cross-entropy loss of discriminator, and thus the gradients are backpropated from discriminator's cross-entropy loss through neural-network-based embedding simulator for updating generator's parameters.
UT-GAN (U-Net and Double-Tanh Framework using GAN) \cite{UTGAN} has improved ASDL-GAN by utilizing more advanced architecture for generator and simulator.
Although UT-GAN has outperformed traditional methods, the neural-network-based embedding simulator would inevitably lead to the dilemma between gradient vanishing and large modification deviation.
SPAR-RL (Steganographic Pixel-wise Actions and Rewards with Reinforcement Learning) \cite{SPARRL} is a new automatic cost learning framework.
With the mechanism of non-differentiable computation in RL, it can overcome the drawbacks of neural-network-based embedding simulators in ASDL-GAN/UT-GAN.
Specifically, it utilizes the policy network to generate embedding policies, i.e., embedding change probabilities, and then utilizes a sampling-based embedding simulator to sample actions, i.e., modifications.
The policy network aims to maximize the rewards assigned for the sampled actions, and thus there is no requirement for propagating gradients through embedding simulator.
In this way, SPAR-RL can simultaneously sample discrete modifications and avoid gradient vanishing issue. SPAR-RL and ASDL-GAN/UT-GAN can share the network structure for policy network and generator, environment network and discriminator, and can share the alternant updating strategy for these two networks. Their differences lie in the structure of embedding simulator and the optimization goal of policy network and generator.}

\red{As for the stage of adjusting the pre-defined embedding costs, adversarial examples technique plays an important role.
ADV-EMB (Adversarial Embedding) \cite{ADVEMB} asymmetrically adjusts a minimum set of embedding costs according to the gradients of target CNN steganalyzer.
Furthermore, the choice of the proportion of adjusted embedding costs is investigated under the game theoretic framework \cite{Shi}.
ADV-EMB can be combined with minmax strategy, wherein the least detectable stego image for the best CNN steganalyzer is selected at each iteration to train a new target CNN steganalyzer.
Please note that GAN/RL and adversarial examples techniques can improve the security performance of steganographic methods in different stages of cost learning, as shown in Fig. \ref{fig:learn costs}.
In \cite{SPARRL}, experimental results show that the embedding costs learned by SPAR-RL can be adjusted by ADV-EMB for further security improvement. In this paper, we focus on the first stage, which aims to learn embedding costs from scratch.}
\fi

\if

SPAR-RL \cite{SPARRL} is an automatic cost learning framework using deep reinforcement learning, wherein an \textit{agent}, playing the role of the steganographer, aims to learn the optimal \textit{embedding policies} associated with embedding costs, while an \textit{environment} gives reward feedbacks to the feasible actions taken by the agent.
Considering the tremendously high search space for image-level modification actions, in SPAR-RL, the image-level action is decomposed into parallel pixel-wise actions.
In doing so, the policy network takes the cover image $\mathbf X = (x_{i,j})^{H \times W}$ as input, and outputs an intermediate matrix  $\mathbf Q =({q}_{i,j})^{H \times W}$.
A policy matrix $\bm{\Pi} = ({\pi}_{i,j})^{H \times W} $ can be computed from $\mathbf{Q}$, where $\pi_{i,j}$ is the embedding policy of image element $x_{i,j}$, i.e., the probability distribution of possible modification actions. The agent applies stochastic sampling on the embedding policies and takes pixel-wise modification actions.
A modification map $\mathbf{M} = (m_{i,j})^{H \times W}$ is formed and then added to the cover image $\mathbf{X} = (x_{i,j})^{H \times W}$ to obtain a simulated stego image $\mathbf{Y}= (y_{i,j})^{H \times W}$.
On the environment side, an environment network is used to calculate a gradient matrix $\mathbf G = (g_{i,j})^{H \times W}$, wherein $g_{i,j}$ is the gradient of the environment networks's loss function with respect to modification $m_{i,j}$.
A reward matrix $\mathbf R = (r_{i,j})^{H \times W}$ is obtained with $\mathbf M = (m_{i,j})^{H \times W}$ and $\mathbf G = (g_{i,j})^{H \times W}$ where the pixel-wise reward $r_{i,j}$ is positive when the signs of $m_{i,j}$ and $g_{i,j}$ are the same, indicating that such modification action should be encouraged. Otherwise the modification action is discouraged. The overall reward is summed up by the pixel-wise rewards.

To learn the optimal embedding policies, the policy network and the environment network are alternately updated in such a way that the former aims at maximizing the overall reward for the taken actions while the latter tries to distinguish between the cover and the simulated stego image and then returns gradients for reward assignment.
When the learning process converges, the embedding policies, i.e., embedding change probabilities, can be inversely converted to embedding costs for practical message embedding.
\fi

\section{Ineffectiveness of applying SPAR-RL to learn JPEG embedding costs}
\label{sec:SPARRLv1v2}

\subsection{Brief Review of SPAR-RL}
\label{sec:SPARRL}
In SPAR-RL \cite{SPARRL}, an {agent} playing the role of the steganographer, aims to learn the optimal {embedding policies} associated with embedding costs, while an {environment}  gives reward feedbacks to the feasible actions taken by the agent.
The actions are simulated embedding modifications stochastically sampled according to the probabilities defined by embedding policies.
Considering the tremendously high search space for image-level modification actions, the image-level action is decomposed into parallel pixel-wise actions in SPAR-RL.

A policy network is responsible for taking the cover image as input, and outputting a policy matrix defining the probability distribution of possible modification actions.
On the environment side, an environment network is used to obtain a reward matrix
to indicate whether the corresponding sampled modification actions should be
encouraged or discouraged.
SPAR-RL-v1 and SPAR-RL-v2 were two implementations proposed in \cite{SPARRL} with different network capacities in policy network and environment network .


To learn the optimal embedding policies, the policy network and the environment network are alternately updated in a similar way as GAN for that the former aims to maximize the overall reward for the taken actions while the latter returns rewards and at the same time updates itself {towards better classification} between the cover and the simulated stego images.
When the learning process converges, the embedding policies can be inversely converted to embedding costs for practical message embedding.

\begin{table}[t!]
\renewcommand\arraystretch{1}
{\caption{
$P_{\text E}$ of steganographic methods against GFR steganalyzer under the setting of JPEG quality factor 75.}
\label{tab:SPARRLv2}}
\centering
\begin{tabular}{ccccccc}
\toprule
\multirow{2}{*}{\textbf{Steganalyzer}}& \textbf{Steganographic}& \multicolumn{5}{c}{\textbf{Payload (bpnzAC)}}  \\\cline{3-6}
& \textbf{method}& {\textbf{0.1}} &  {\textbf{0.3}}&  {\textbf{0.5}} \\\midrule
\multirow{2}{*}{GFR}&J-UNIWARD  & 45.38\%  & 29.56\% & 15.01\%\\
&A-SPAR-RL-v2 &42.95\% &29.20\% &13.97\% \\

\bottomrule
\end{tabular}
\vspace{0.2cm}
\renewcommand\arraystretch{1}
{\caption{
$P_{\text E}$ of A-SPAR-RL-v2 and its environment network in five independent experiments.}
\label{tab:5experiment}}
\centering
\begin{tabular}{p{2.5cm}p{0.7cm}<{\centering}p{0.7cm}<{\centering}p{0.7cm}<{\centering}p{0.7cm}<{\centering}p{0.6cm}<{\centering}}
\toprule
 &$\#1$ &$\#2$ &$\#3$ &$\#4$ &$\#5$ \\\midrule
A-SPAR-RL-v2 & 11.62\% & 11.56\% &21.72\% &22.04\% &12.37\% \\
Environment network & 44.75\% & 42.80\% &28.04\% &37.45\% &40.04\% \\
\bottomrule
\end{tabular}
\end{table}

\begin{figure}[t!]
\centering
\includegraphics[width=0.45\textwidth]{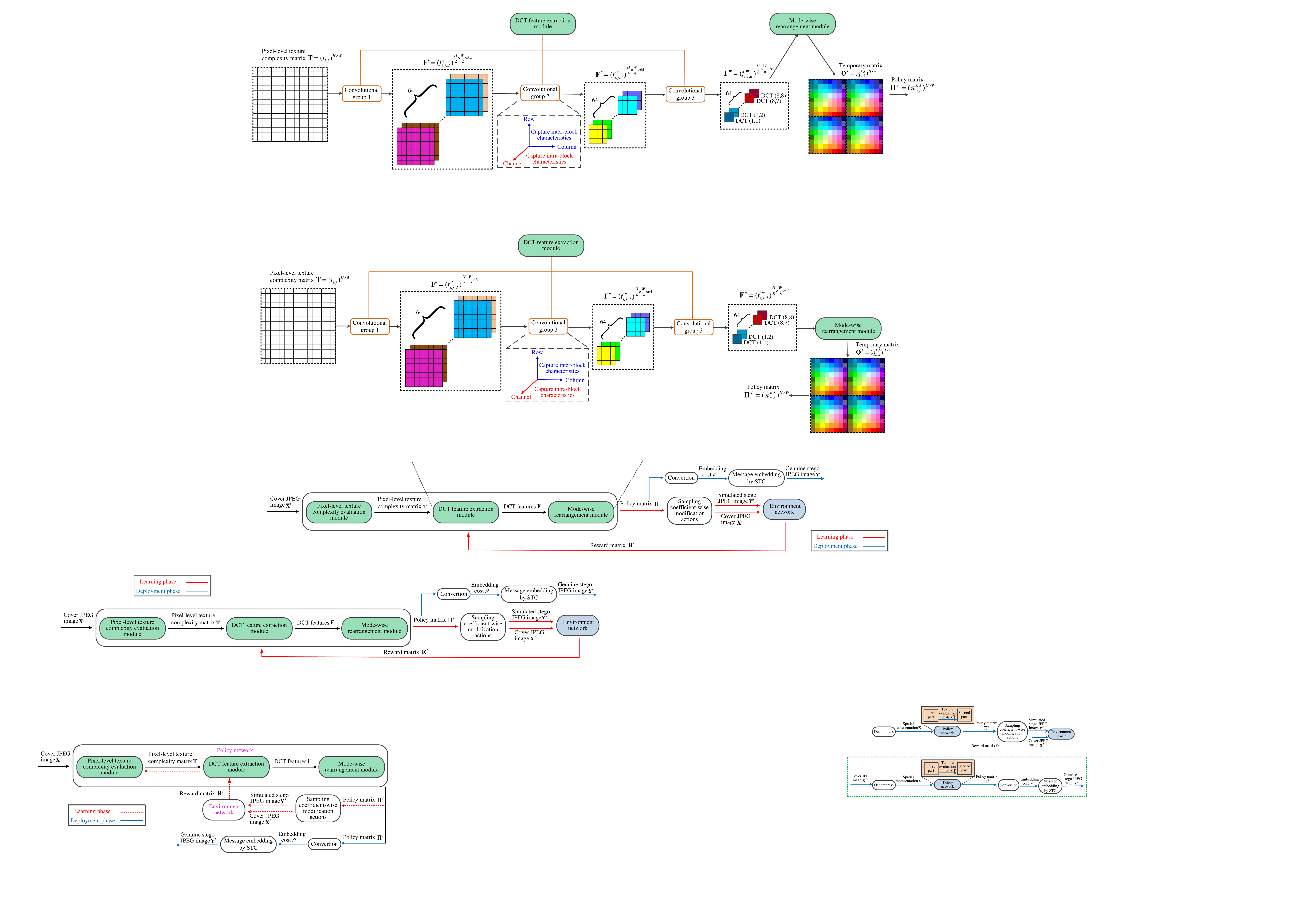}
\caption{The overview diagram of the proposed JEC-RL.}
\label{fig:diagram}
\vspace{-0.8cm}
\end{figure}

\subsection{Ineffectiveness of SPAR-RL for JPEG}
\label{sec:disSPARRL}

SPAR-RL cannot be well applied on JPEG images.
The reason comes from two aspects.
First, the environment networks are designed to capture the embedding traces left in spatial domain and may fail to evaluate discriminative features in DCT domain. Therefore, the returned rewards may not provide effective evaluation for DCT coefficients' modification actions on 64 DCT modes. Besides, such environment network architecture may not be able to converge steadily in the situation of JPEG cost learning.
Second, the policy network works on spatial pixels rather than DCT coefficients. In contrast to spatial pixels, DCT coefficients are arranged into $8\times8$ frequency-mode structure and thus exhibit both inter-block and intra-block correlations
Both correlations have impacts on the security performance, but may not be fully captured by the common convolution operations used in SPAR-RL.

We have conducted an experiment based on an adapted version of SPAR-RL-v2, abbreviated as A-SPAR-RL-v2, wherein the JPEG images are decompressed to spatial domain before fed into the policy network and environment network. The policy network directly outputs embedding policies for DCT coefficients.
Experiments were conducted on the $\textit{BOSSBase}$ with JPEG quality factor 75, and GFR was used for evaluation. Detailed experimental settings were the same as those given in Section \ref{sec:setting}.
The results are shown in Table \ref{tab:SPARRLv2}. We can observe that although A-SPAR-RL-v2 can still work to some extent, its performance is inferior to a baseline method J-UNIWARD, not to mention other methods proposed recently.

Besides, A-SPAR-RL-v2 would encounter severe convergence issue.
As shown in Table \ref{tab:5experiment}, among five independent experiments, the security performance of A-SPAR-RL-v2 ranges from $11.56\%$ to $22.04\%$ in the case of 0.4 bpnzAC.
To further investigate the phenomenon, we use the environment network in A-SPAR-RL-v2 to classify the cover images and stego images generated by J-UNIWARD on 1.0 bpnzAC.
We can observe that the model whose environment network has better detection performance is more likely to learn a more secure cost learning method.
However, the environment network in A-SPAR-RL-v2 has not considered the JPEG domain knowledge, and thus its detection performance is quite unstable.
To implement effective JPEG cost learning method, the mechanism of the environment network should be further investigated.

\section{JEC-RL by exploiting JPEG Domain knowledge}
\label{sec:method}



In order to address the issue of automatic cost learning for JPEG images, we propose JEC-RL based on JPEG domain knowledge.
We first give an overview of the proposed method, and then
show the details of two important components, i.e., the policy
network following the domain-transition design paradigm and the gradient-oriented environment network.

\subsection{Method Overview}
The design methodology of JEC-RL is extended from SPAR-RL framework \cite{SPARRL}, where a policy network yields optimal embedding policies through iterative interactions with an environment network, as shown in Fig. \ref{fig:diagram}.
In this paper, the policy network follows a domain-transition paradigm consisting of three modules.
The first module takes JPEG image elements $\mathbf X^{J} = (x_{a,b}^{k,l})^{H \times W }$ as input and outputs a matrix $\mathbf T = (t_{i,j})^{H \times W }$, which contains the evaluation of texture complexity for each pixel in spatial domain.
The second module transforms spatial texture complexity features $\mathbf T = (t_{i,j})^{H \times W }$ to DCT frequency features $\mathbf F = (f_{i,j,d})^{H/8 \times W/8 \times 64}$ via spatial aggregation and frequency conversion.
The third module rearranges the DCT features into $8\times8$ DCT mode structure
{to obtain a policy matrix} $\bm{\Pi}^{J} = (\bm{\pi}_{a,b}^{k,l}(m))^{H \times W }$, where $\bm{\pi}_{a,b}^{k,l}(m)$ is the embedding policy, i.e., the probability of possible modification actions, for image element $x_{a,b}^{k,l}$.
In ternary embedding, the possible modification actions are $+1, 0, -1$.
Then, the agent can apply stochastic sampling following the embedding policies and take the corresponding sampled pixel-wise modification actions.
A modification map $\mathbf M^{J} = (m_{a,b}^{k,l})^{H \times W }$ is formed and then added to the cover image $\mathbf X^{J} = (x_{a,b}^{k,l})^{H \times W }$ to obtain a simulated stego image $\mathbf Y^{J} = (y_{a,b}^{k,l})^{H \times W }$.
As for the environment network, it is responsible for reward assignment, where the rewards are defined associated with the back-propagated gradients.
In our work, a wide architecture
equipped with a fixed preprocessing layer with $8\times8$ DCT bases is proposed to accomplish the task well.
The environment network takes ${\mathbf X}^{J}$ and ${\mathbf Y}^{J}$ as input, and then calculates a gradient matrix ${\mathbf G}^{J} = (g_{a,b}^{k,l})^{H \times W}$, wherein $g_{a,b}^{k,l}$ is the gradient of the environment network's loss function with respect to the sampled modification $m_{a,b}^{k,l}$.
A reward matrix ${\mathbf R}^{J} = (r_{a,b}^{k,l})^{H \times W}$ is obtained with ${\mathbf M}^{J}$ and ${\mathbf G}^{J}$ as shown in \eqref{eq:DG function}, wherein $r_{a,b}^{k,l}$ evaluates the contribution of the action $m_{a,b}^{k,l}$ on deceiving the steganalyzer.
In the learning phase, the policy network is trained with the environment network by iteratively generating simulated stego images and then receiving rewards for updating its learning parameters. %
In the deployment phase, the well-trained policy network is used to output a policy matrix which can be converted to embedding costs.
With the help of steganography codes such as STC  \cite{STCs}, genuine stego images can be generated.

{
The proposed JEC-RL can take the advantages of the JPEG domain knowledge in the network design.
On the policy network's side, the domain-transition paradigm can capture not only
 the texture level of DCT blocks from spatial domain but also the correlation
among DCT coefficients, while different impacts of DCT
frequency-modes can be implicitly learned through the interactions
with the environmental network.
On the environment network's side, the preprocessing layer with $8\times8$ DCT bases can provide sufficient frequency resolution and is capable of propagating useful gradients for modification actions on different DCT modes. }

\subsection{Domain-transition Paradigm Based Policy Network}
\label{sec:policy network}

\begin{figure*}[t!]
\centering
\includegraphics[width=1.0\textwidth]{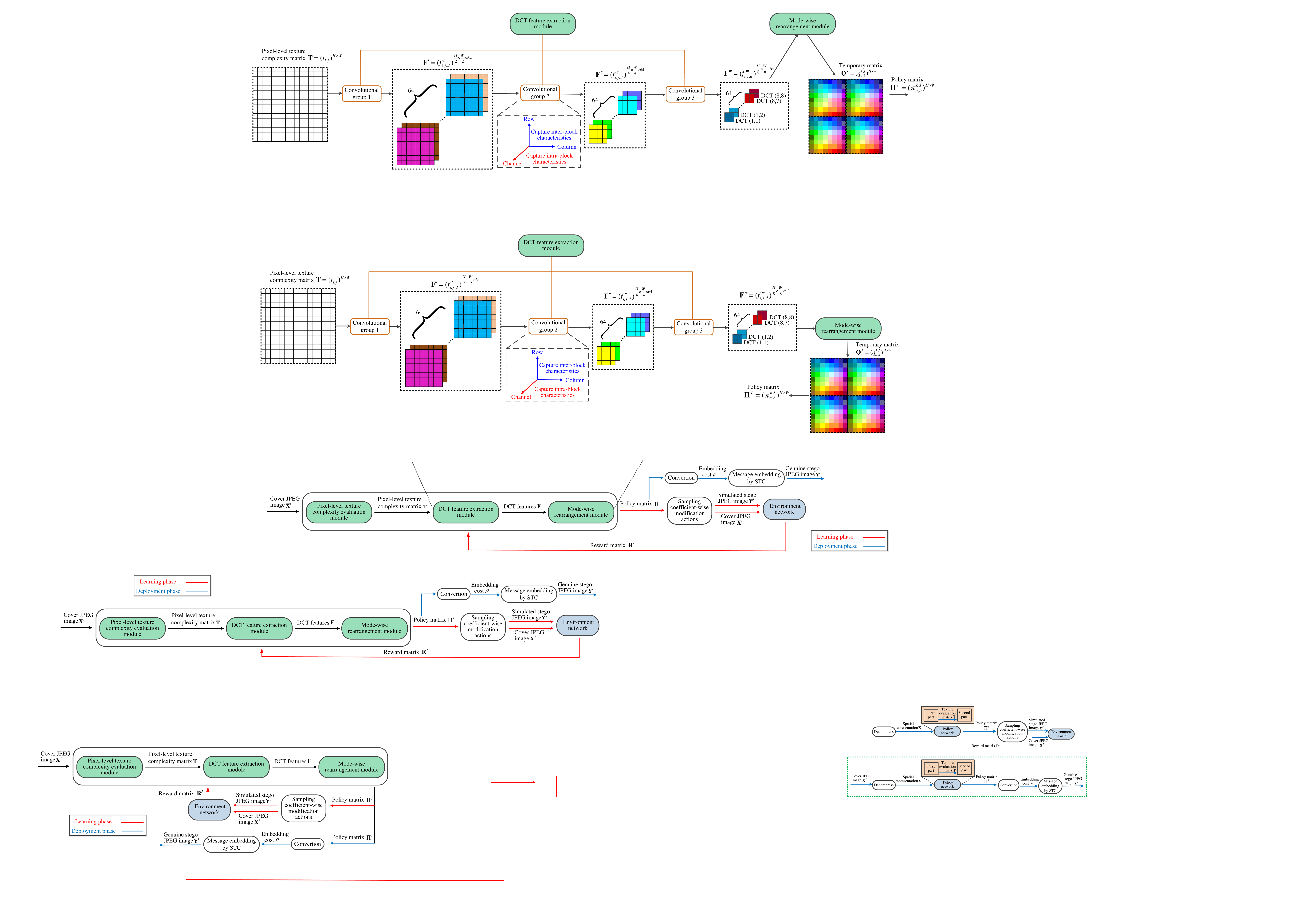}
\caption{
Illustration of the DCT feature extraction and mode-wise rearrangement module.
Each color in the feature map corresponds to a specific frequency.
}\label{fig:ModeWise}
\vspace{-0.5cm}
\end{figure*}

Although it is straightforward to let the policy network operate in DCT domain for reducing computational complexity, it is rather difficult to directly apply convolution operations to DCT coefficients to extract effective features due to the peculiarity of the $8\times8$ DCT mode structure in JPEG images. As a result, in our proposed domain-transition paradigm with three consecutive modules, we firstly evaluate texture complexity in spatial domain, and then extract DCT features via spatial aggregation and frequency conversion, and finally arrange them according to DCT mode structure.

\subsubsection{Pixel-level texture complexity evaluation module}

This module evaluates image texture complexity for each pixel in spatial domain.
In general, such a task can be accomplished by a pixel-to-pixel CNN structure.
{Different network architecture can be applied according to the development of network design in the area of deep learning. In this paper, for fair comparison with SPAR-RL-v2 and JS-GAN, a U-Net \cite{UNet} alike structure is adopted, and its architecture is given in Fig. \ref{fig:fp} in Appendix.}
{On top of the CNN,
a set of inverse DCT (IDCT) basis filters with a stride of $8$
can be used to decompress the JPEG
image $\mathbf X^{J}$ into spatial representation $\mathbf X = (x_{i,j})^{H \times W }$ as the pixel input.}
In order to preserve more information, the elements in $\mathbf X$ are of floating values without rounding.
This first module outputs a matrix $\mathbf T = (t_{i,j})^{H \times W }$, where $t_{i,j}$ denotes the texture complexity of pixel $x_{i,j}$.

{In fact, it is possible to use the texture calculation process in existing JPEG steganographic methods for obtaining the texture complexity evaluation information.
For example, the image residuals obtained with three Daubechies wavelet filters in J-UNIWARD can be used as a pixel-level texture complexity matrix $\mathbf T = (t_{i,j,d})^{H \times W \times 3}$.
For another example, an 8 times
nearest-neighbor upsampling can be performed on the block energy
in UERD and given in \eqref{eq:E} to yield a pixel-level texture complexity matrix $\mathbf T = (t_{i,j})^{H \times W }$.}
{
Applying such texture calculation process into this module in our proposed cost learning method
can be considered as a way to improve existing methods, because the two subsequent modules in the policy network can be used afterwards to obtain better costs.
But it may be inferior to the proposed cost learning method whose first module is implemented as a pixel-to-pixel CNN,  because the texture calculation process is not learnable.
}

\subsubsection{DCT feature extraction module}

This module
takes pixel-level texture complexity matrix 
as input, and then aggregates pixel-level texture features in spatial domain and converts them into frequency features as $\mathbf{F}^{\prime\prime\prime} = ({f}^{\prime\prime\prime}_{i,j,d})^{H/8 \times W/8 \times 64}$, wherein ${f}^{\prime\prime\prime}_{i,j,d}$ is the DCT feature corresponding to the DCT coefficient in the $(i,j)$-th
DCT block and
$(\lceil d/8\rceil,(d-1)\%8+1)$-th
DCT mode.
To realize this function, the second module can be composed of a stack of three convolutional groups,
as shown in Fig. \ref{fig:sp} and illustrated detailedly in Fig. \ref{fig:ModeWise}.
The number of convolutional kernels in each group is set to be $64$, which equals to the total number of JPEG DCT frequency-modes.
In each convolutional group, the row and the column dimensions of the output feature maps are reduced by a half with respect to the input feature maps via performing convolution with a stride of $2$.
In this way, a 2-D input $\mathbf T =(t_{i,j})^{H \times W}$ can be turned into a 3-D feature map $\mathbf{F^{\prime\prime\prime}} = ({f}^{\prime\prime\prime}_{i,j,d})^{H/8 \times W/8 \times 64}$.
Note that a receptive field of $15\times15$ spatial texture complexity features is aggregated and then converted into a $1\times1\times64$ frequency feature vector. As shown in Fig. \ref{fig:ModeWise},
in all three convolutional groups, the intra-block characteristics can be captured via convolution along the channel direction,
while the inter-block characteristics can be captured via convolution along the column and the row direction. The channel dimension of the feature maps outputted by the last convolutional layer equals to the number of $64$ DCT frequency-modes, and each feature map is associated with a specific DCT frequency-mode. The extracted DCT features are related to coefficient-wise embedding policies.
Hence, to restrict them as a kind of probabilities, we
use a Sigmoid activation function at the end of the last convolutional group
to make the values of ${f}^{\prime\prime\prime}_{i,j,d}$ be in the range of $[0,1]$.

It should be noticed that this module performs the transition of features from spatial domain to DCT domain. But it is different from performing a DCT transform.
Firstly, the transition is not designed to be reversible like DCT, because the goal here is to
obtain embedding policies instead of performing frequency analysis.
Secondly, the receptive field may not be restricted to be $8\times8$, but depends on the implemented network structure.
In our implementation, the receptive field is $15\times15$.

\subsubsection{Mode-wise rearrangement module}
A mode-wise rearrangement module is performed at the end of the policy network to rearrange the 3-D feature maps $\mathbf{F^{\prime\prime\prime}} = ({f}^{\prime\prime\prime}_{i,j,d})^{H/8 \times W/8 \times 64}$ into a 2-D temporary matrix $\mathbf Q^{J} =(q_{a,b}^{k,l})^{H \times W}$
as
\begin{equation}\label{}
  {q}_{a,b}^{k,l} = {f}^{\prime\prime\prime}_{a,b,(k-1)*8+l}.
\end{equation}
In this way, the $((k-1)*8+l)$-th feature map in $\mathbf{F}$ is forced to be linked with the $(k,l)$-th DCT frequency-mode for learning embedding policy, while the spatial positions of DCT blocks are preserved.

The final outputted policy tensor $\bm{\Pi}^{J} = (\bm{\pi}_{a,b}^{k,l}(m))^{H \times W }$ can be obtained by:
    \begin{equation}
     \begin{aligned}
        \bm{\pi}_{a,b}^{k,l}(m=1)=\bm{\pi}_{a,b}^{k,l}(m=-1)
    =q_{a,b}^{k,l}/2,
     \label{equ:q1}
     \end{aligned}
     \end{equation}
     \begin{equation}
     \begin{aligned}
        \bm{\pi}_{a,b}^{k,l}(m=0)=1-q_{a,b}^{k,l},
     \label{equ:q2}
     \end{aligned}
     \end{equation}
where $m \in \mathcal{M} = \{-1, 0, 1\}$ is the possible modification actions in ternary embedding.

{
Our proposed mode-wise rearrangement is an inverse process of the $8\times8$ phase split module
proposed in PNet \cite{PNet}, which splits a feature map into 64 sub-feature maps in order to form individual phase-aware branches for JPEG steganalysis.
These two modules are founded on the same JPEG domain knowledge but their roles are different.
The phase split module is used for decomposing features
and such a procedure can be found in conventional steganalysis or forensics.
By contrast, our proposed mode-wise rearrangement module is used for composing features,
and to the best of our knowledge, it is introduced for the first time.}

\subsubsection{Learning procedure}
\label{Sec:Opt}
In the learning phase, the modification actions in ${\mathbf{M}}^{J} = ({m}_{a,b}^{k,l})^{H \times W }$ are sampled in a coefficient-wise manner according to
the policy matrix $\bm{\Pi}^{J} = (\bm{\pi}_{a,b}^{k,l}(m))^{H \times W }$.
Then, the environment network is used to evaluate the contribution of modification actions and returns a reward matrix ${\mathbf{R}}^{J} = ({r}_{a,b}^{k,l})^{H \times W }$, as described in detail in the next sub-section.
{In general, when a larger reward is assigned for ${m}_{a,b}^{k,l}$, via maximizing the overall rewards,
a larger probability $\pi_{a,b}^{k,l}(m={m}_{a,b}^{k,l}|\mathbf{X}^{J})$ would be obtained for the sampled modification
in the next training round, and vice versa.}
The parameters of the policy network can be updated with a loss function ${l_{ A}}$ defined by the weighted summation of a reward loss ${l_{ R}}$ and a capacity loss ${l_{ C}}$ as follows:
\begin{equation}
   {l_{ A}} =  \alpha \cdot {l_{ R}} + \beta \cdot   {l_{ C}},
   \label{eq:GenLoss}
\end{equation}
\begin{equation}
   \small {l_{ R}} = -\frac{1}{H \times W}\sum_{a=1}^{H/8}\sum_{b=1}^{W/8}\sum_{k=1}^{8}\sum_{l=1}^{8}
   r_{a,b}^{k,l} \cdot \text{log} \pi_{a,b}^{k,l}(m={m}_{a,b}^{k,l}|\mathbf{X}^{J}),
   \label{eq:reward loss}
\end{equation}
\begin{equation}
\begin{aligned}
  \small {l_{ C}} = & \left(-\sum_{a=1}^{H/8}\sum_{b=1}^{W/8}\sum_{k=1}^{8}\sum_{l=1}^{8} \sum_{m\in\mathcal{M}} \right.\\
   &\left.\pi_{a,b}^{k,l}(m|\mathbf{X}^{J})\text{log}_{2}\pi_{a,b}^{k,l}(m|\mathbf{X}^{J})-C \right)^{2},
   \label{eq:capacity loss}
\end{aligned}
\end{equation}
where $\alpha$ and $\beta$ are the weights, and $C$ is the target capacity.
The derivation of \eqref{eq:reward loss} follows the RL procedure \cite{PolicyGradient} and the details
can also be referred to \cite{SPARRL}, and
the capacity loss \eqref{eq:capacity loss} is originated from \eqref{equ:capacity}.

In the deployment phase, embedding costs can be converted from the policy tensor outputted by the well-trained policy network as
\begin{equation}
  \label{eq:p_to_rho}
  \rho_{a,b}^{k,l}=\ln \left(\frac{2}{q_{a,b}^{k,l}}-2 \right).
\end{equation}

\begin{figure}[t!]
\centering
\subfigure[$8\times8$ DCT basis filters]{\includegraphics[height=2.4cm]{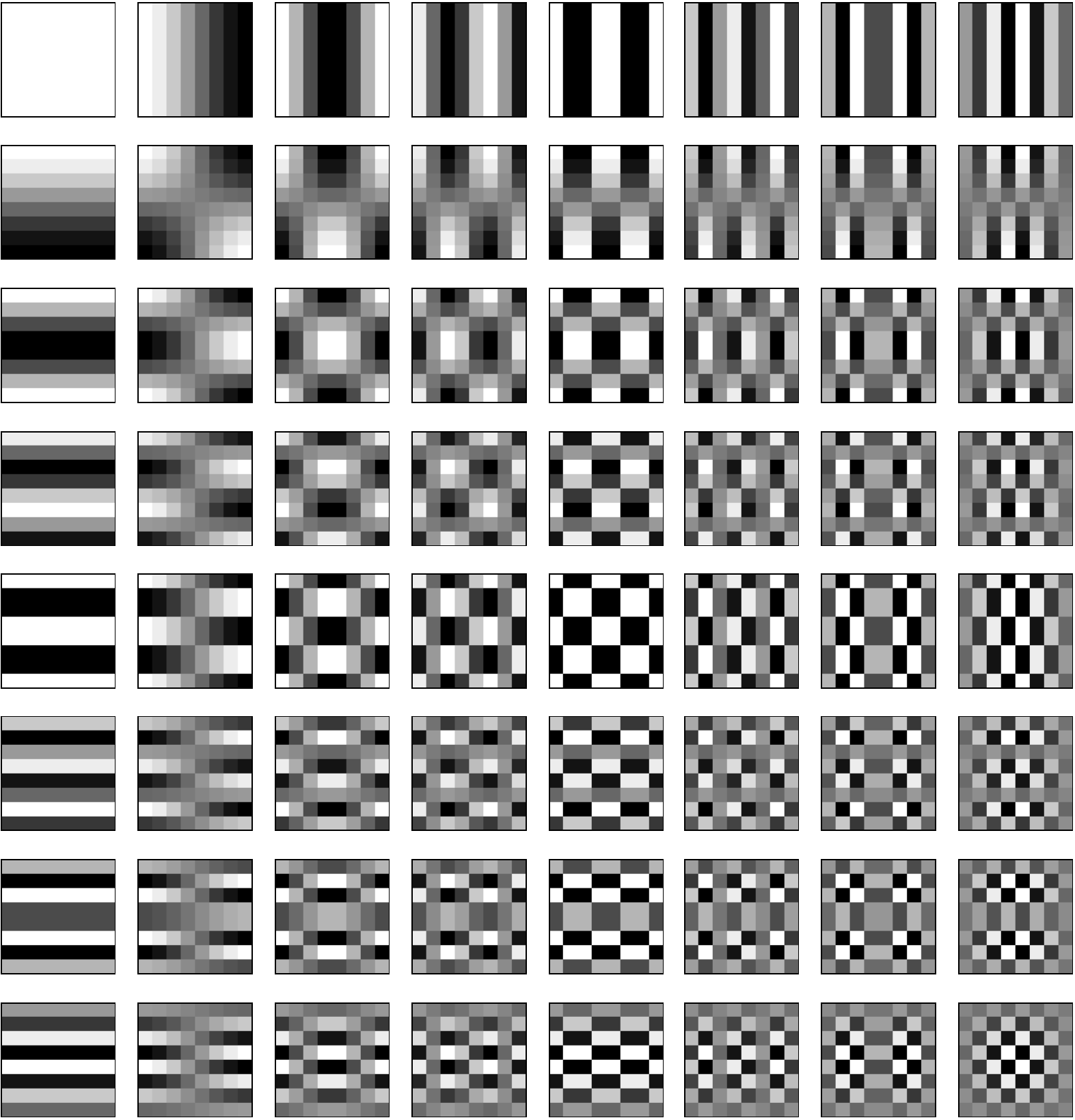}
\label{fig:64DCT_s}}
\hspace{.05in}
\subfigure[$4\times4$ DCT basis filters ]{\includegraphics[height=2.4cm]{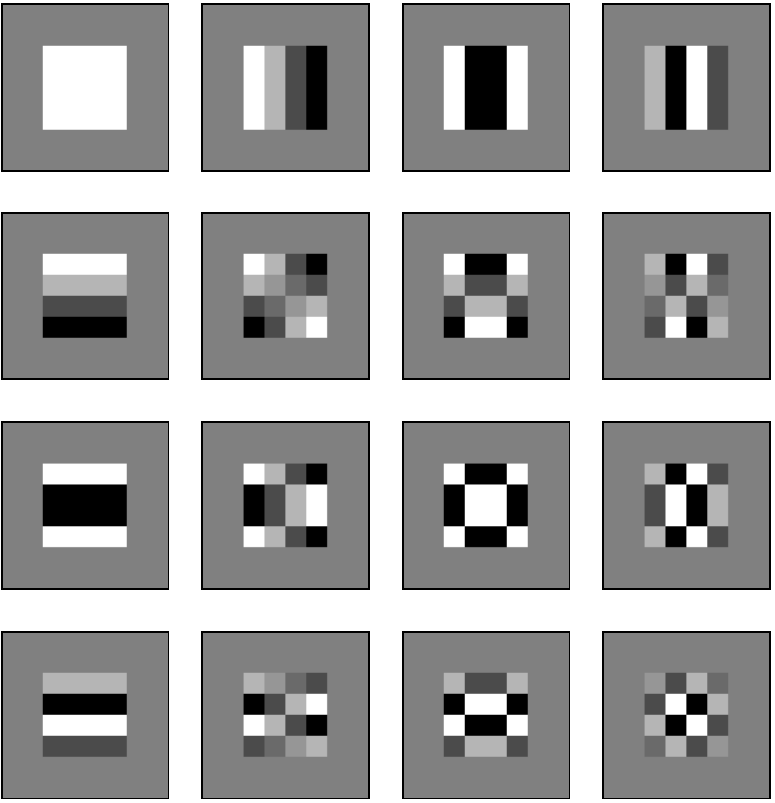}
\label{fig:16DCT_s}
}
\hspace{.05in}
\subfigure[SRM high-pass filters ]{\includegraphics[height=2.4cm]{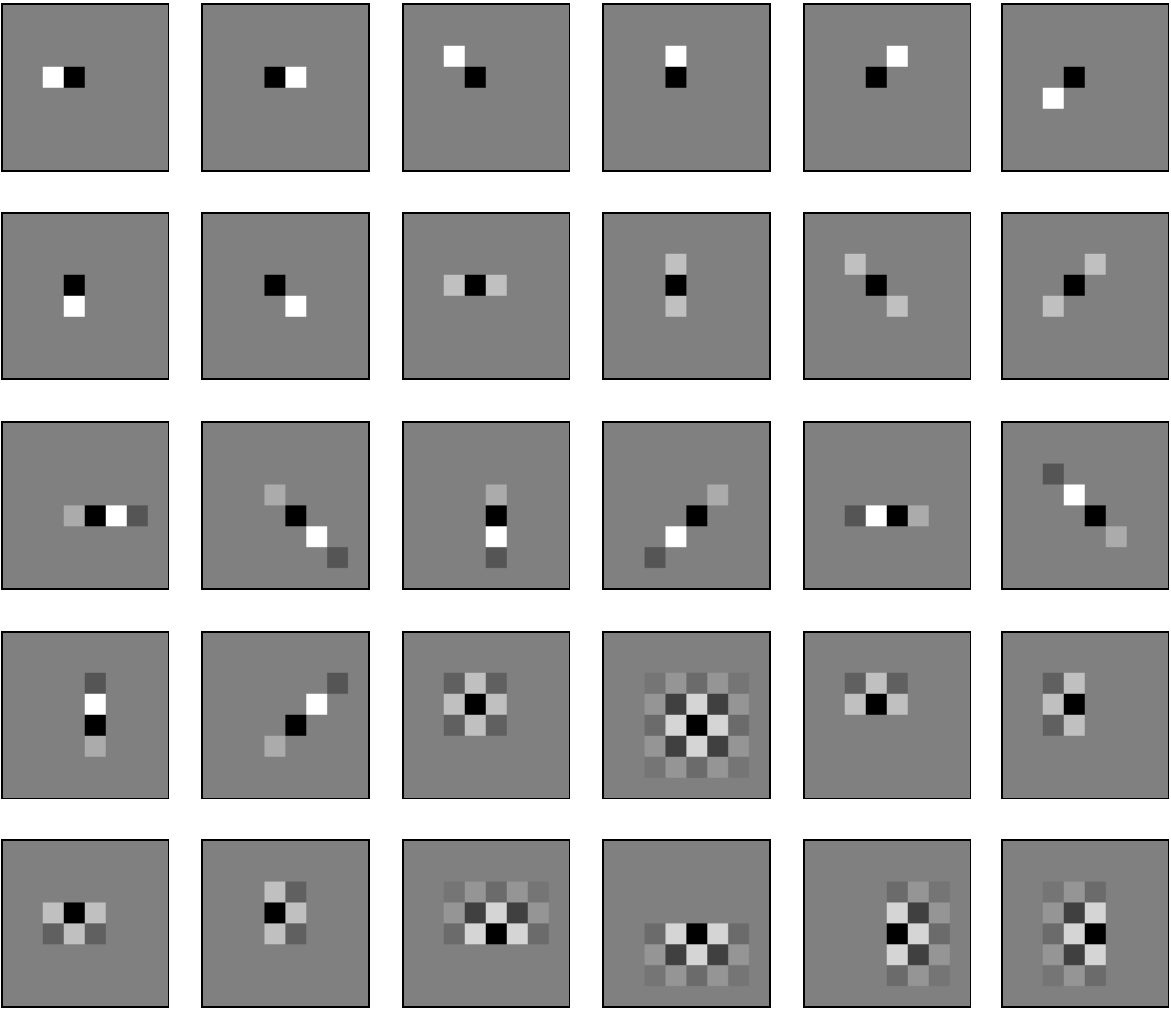}
\label{fig:30SRM_s}
}
\hspace{.05in}
\subfigure[DCT of $8\times8$ DCT basis filters ]{\includegraphics[height=2.4cm]{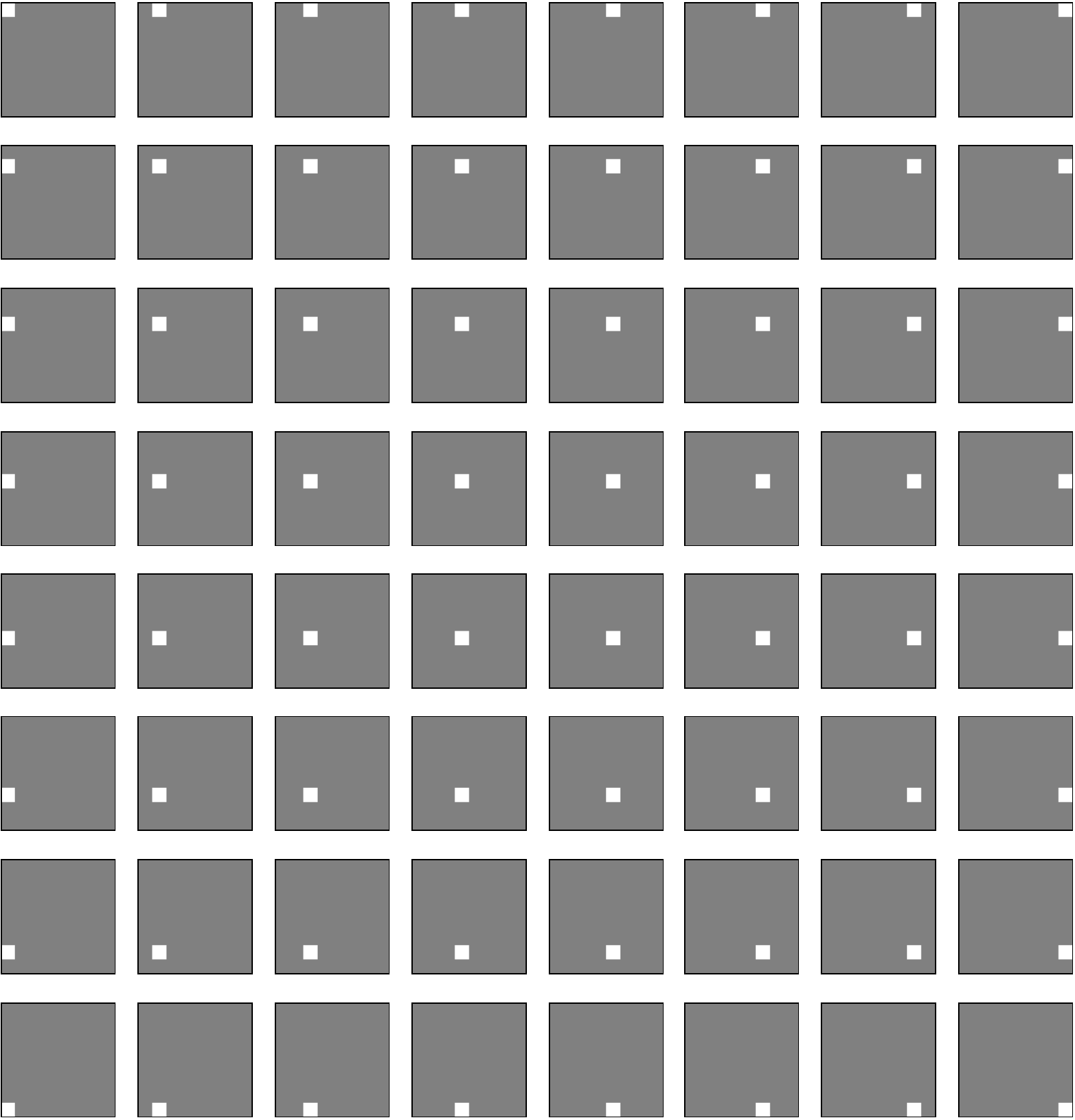}
\label{fig:64DCT_f}
}
\hspace{.05in}
\subfigure[DCT of $4\times4$ DCT basis filters ]{\includegraphics[height=2.4cm]{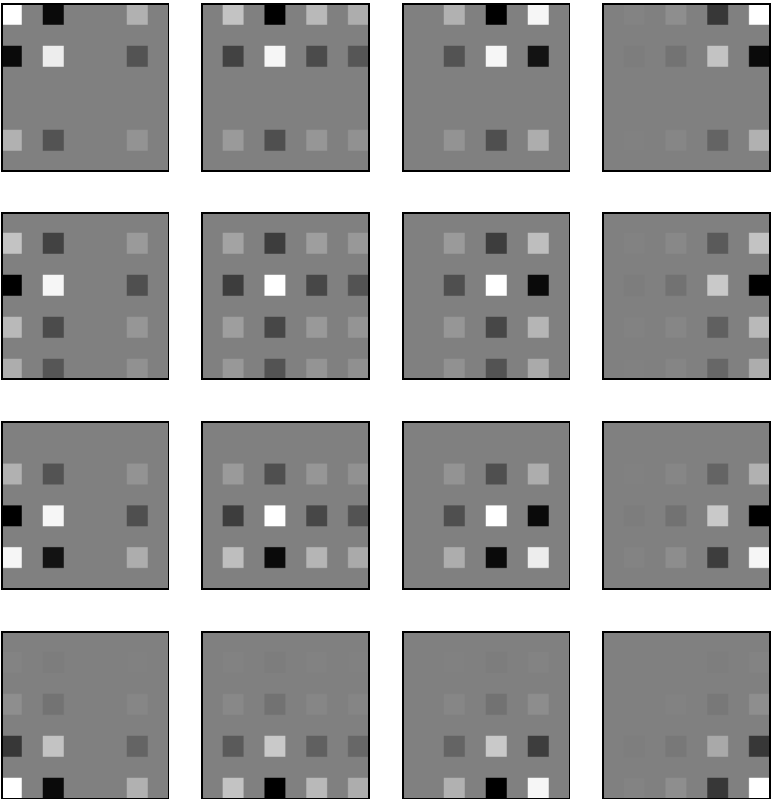}
\label{fig:16DCT_f}
}
\hspace{.05in}
\subfigure[DCT of SRM high-pass filters]{\includegraphics[height=2.4cm]{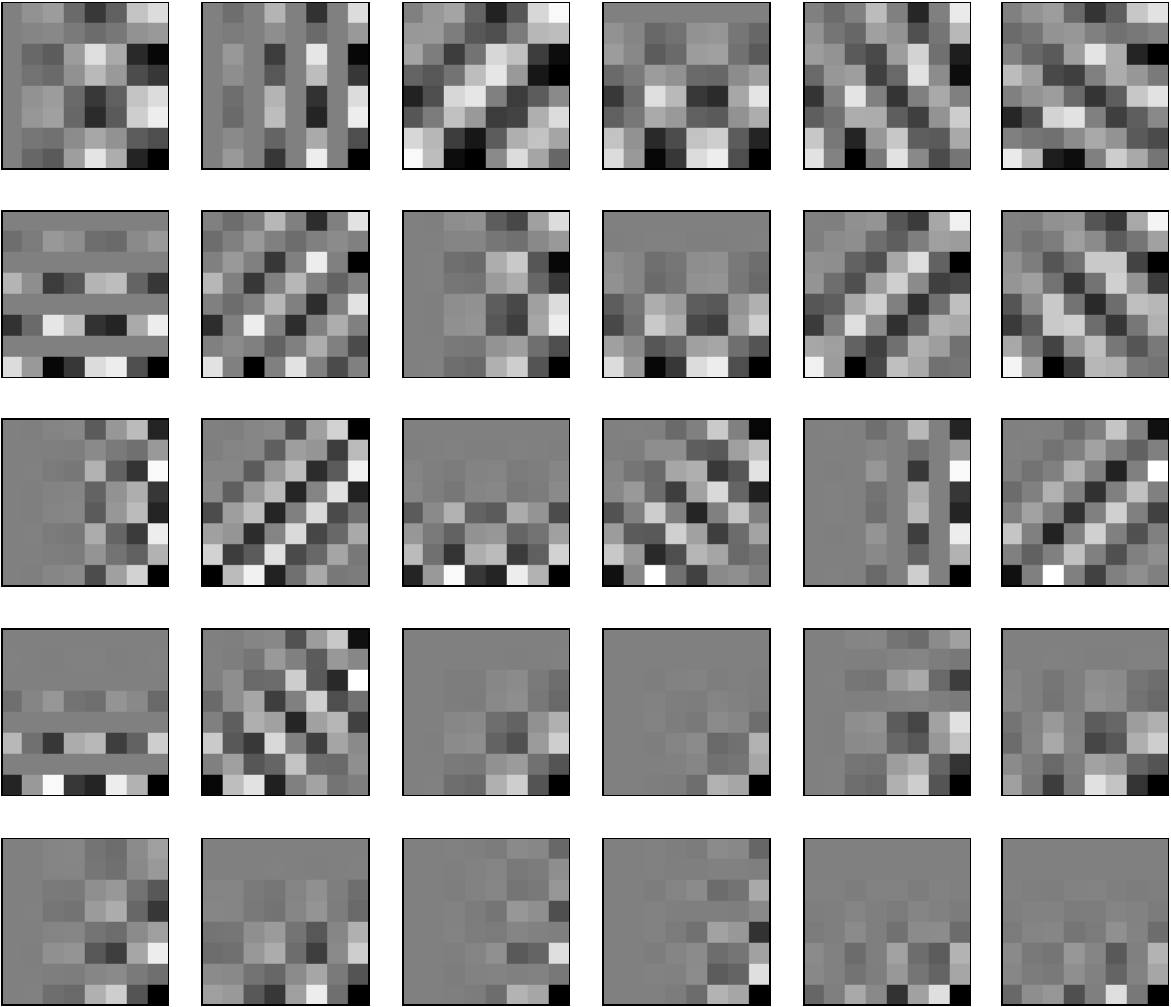}
\label{fig:30SRM_f}
}
 {\caption{Illustration of the DCT transform of different filter sets.}
 \label{fig:DCTtransform}}
 \vspace{-0.6cm}
\end{figure}

\subsection{Gradient-oriented Environment Network }
\label{sec:environment network}

As shown in \eqref{eq:reward loss}, it is important for the environment to provide discriminative rewards in the training process of JEC-RL.
Following \cite{SPARRL}, the coefficient-wise reward value in the reward matrix $\mathbf R^{J} = (r_{a,b}^{k,l})^{H \times W}$ is computed by
\begin{equation}\label{eq:DG function}
\begin{aligned}
r_{a,b}^{k,l} =\xi \cdot \text{sign}({m}_{a,b}^{k,l}) \cdot g_{a,b}^{k,l},
\end{aligned}
\end{equation}
where $\xi$ is a constant,
${m}_{a,b}^{k,l}$ is the sampled modification,
and $g_{a,b}^{k,l}$ is the gradient of the environment network's loss function $l_{ E}$ with respect to modification ${m}_{a,b}^{k,l}$.
The cross-entropy loss function of the environment network is
defined as
   \begin{equation}
    \begin{aligned}
    {l_{ E}} =  -z'_{0}log(z_{0})-z'_{1}log(z_{1}),
    \label{eq:DisLoss}
    \end{aligned}
    \end{equation}
where $z_{0}$ and $z_{1}$ are the environment network's Softmax outputs for the cover image $\mathbf X^{J}$ and the simulated stego images ${\mathbf Y}^{J}$, respectively, and $z'_{0}$ and $z'_{1}$ are their corresponding ground-truth labels.
The environment network is not used during the deployment phase.

{The reward $r_{a,b}^{k,l}$ is positive when the signs of $m_{a,b}^{k,l}$ and $g_{a,b}^{k,l}$ are the same, indicating that such modification action should be encouraged. Otherwise the modification action is discouraged.}
{Please note that in a general RL setting, the agent aims at maximizing the overall reward, and it is not necessary to compete against a steganalyzer.
However, in the case of JEC-RL, the reward is constructed by the environment network's gradient, and therefore maximizing such reward is more or less related to compete against the environment network.}
Compared with a CNN steganalyzer which aims at achieving better detection accuracy, an environment network focuses more on providing informative rewards to improve the performance of the policy network.
For such a purpose, rather than using an off-the-shelf JPEG steganalyzer, we propose a gradient-oriented environment network modified from XuNet \cite{Xu} via utilizing an effective preprocessing layer and a wide network architecture. Its architecture is given in Fig. \ref{fig:e}

\subsubsection{Preprocessing layer with $8\times8$ DCT basis filters}

To capture embedding traces in both spatial and DCT domain while providing coefficient-wise rewards,
we use a JPEG-tailored preprocessing layer to handle the image.
A bank of  $8\times8$ DCT basis filters is used in the preprocessing layer to obtain image residuals,
where each DCT basis filter is denoted as $\mathbf Z^{u,v} = (z_{i,j}^{u,v})^{8 \times 8}$ ($1 \leq i,j,u,v \leq 8$), and its element is computed as
\begin{equation}
  \label{eq:z}
  z_{i,j}^{u,v} = \frac{{w}_{u}{w}_{v}}{8}\text{cos}(\frac{\pi u(2i+1)}{16})\text{cos}(\frac{\pi v(2j+1)}{16}),
\end{equation}
where ${w}_{0}=1$, ${w}_{k}=\sqrt{2}$ for $k>0$,
$(u,v)$ denotes the frequency index of the filter,
and $(i,j)$ denotes the position of the element in the matrix.
To this end, in preprocessing layer, we decompress a JPEG image $\mathbf X^{J}$  into spatial representation without rounding $\mathbf X$ as input, and then obtain image residuals by $8\times8$ DCT basis filters.
Finally, to facilitate effective learning, we limit the range of the image residuals to $[-8,8]$ by ReLU-based truncation.

We note that the JPEG CNN steganalyzer J-XuNet and Zeng-CNN obtain their best steganalytic detection performance with a non-learnable preprocessing layer implemented by $4\times4$ and $5\times5$ DCT basis filters, respectively.
SRNet  abandons a fixed preprocessing layer and utilizes multiple stacked learnable convolutional layers to capture image residuals.
As for spatial CNN steganalyzer, XuNet utilizes preprocessing layer equipped with SRM high-pass filters.
Although these preprocessing layers have obtained satisfied performance in detection task, the case for JPEG cost learning is quite different.
Since the modification actions are possible to be taken place in all $8\times8$ DCT modes,
the preprocessing layer in the environment network should provide sufficient frequency resolution to cover all 64 frequency bands and provide discriminative rewards for all DCT modes  as well.

{We apply DCT transform to $8\times8$ DCT basis filters, $4\times4$ DCT basis filters, and 30 SRM high-pass filters, respectively, to investigate their frequency responses.
Filters are zero-padded to $8\times8$ if necessary.
The results are shown in Fig. \ref{fig:DCTtransform}.
There is no doubt that each filter in the $8\times8$ DCT basis filter set responses to a specific DCT frequency.
For the other two filter sets, their frequency responses mainly focus on some particular frequency bands and fail to fully cover all 64 frequency bands.
In fact, covering a wide range of frequency band in the environment network's preprocessing layer is beneficial for propagating effective gradients to modification actions on different DCT modes.
We will give experimental evidences in Section \ref{sec:investigation}.}

\subsubsection{Learnable layers with wide network structure}

As for the learnable layers in gradient-oriented environment network, the 5-layer XuNet is used as backbone.
To further improve the network capacity, we can increase the network width or network depth.
Note that although a CNN steganalyzer with a deeper structure, such as 22-layer J-XuNet or 12-layer SRNet, can achieve better steganalytic performance, the case for the environment network is different.
The environment network aims at providing more informative gradient information, and the deep architecture would be inefficient and unstable to propagate informative gradients for DCT coefficients.
Therefore, we use the 5-layer XuNet as backbone, and expand the width of each layer.
{Specifically, the width of the first to the fifth layer is expanded from 8, 16, 32, 64, 128 to 48, 48, 64, 128, 256, respectively.}
At the end of the network, there is a fully-connected layer followed by a Softmax function.

\section{Experiments}
\label{experiment}

In order to evaluate the performance of the proposed JEC-RL,
extensive experiments were carried out, including evaluating the security performance against state-of-the-art steganalyzers (given in Section \ref{sec:performance}),
conducting ablation studies on network architecture design (given in Section \ref{sec:ablation}),
incorporating the texture calculation process of traditional methods into design paradigm (given in Section \ref{sec:refine}),
investigating the effectiveness of the $8\times8$ DCT basis filters in environment network (given in Section \ref{sec:investigation}), and
visualizing the embedding pattern (given in Section \ref{sec:visulazation}).
The experimental settings are described in Section \ref{sec:setting}.

\begin{table*}[t!]
\renewcommand\arraystretch{1}
\scriptsize
{
\caption{$P_{\text E}$ of steganographic methods against different steganalyzers under the setting of JPEG quality factor 75.}
\label{tab:basic500k}}
\centering
\begin{tabular}{cclllll}
\toprule
\textbf{Steganalyzer} &
\textbf{Steganographic method}&
\textbf{0.1 bpnzAC} & \textbf{0.2 bpnzAC} & \textbf{0.3 bpnzAC} & \textbf{0.4 bpnzAC} & \textbf{0.5 bpnzAC}
\\\midrule

\multirow{4}{*}{PHARM}
&J-UNIWARD  & 46.44\%  & 39.98\% & 32.42\% & 24.46\% & 17.54\%\\
\cdashline{2-7}
&J-MSUNIWARD  & 46.90\%& 40.86\% & 33.44\%& 26.15\% & 19.19\%\\
&MSUERD\_SPA & 46.45\%& 40.69\% & 34.52\%& 27.85\% & 21.92\%\\\cdashline{2-7}
&\textbf{JEC-RL} & \textbf{47.36\%} & \textbf{43.19\%} & \textbf{38.13\%} & \textbf{32.24\%} & \textbf{26.12\%}\\\midrule

\multirow{4}{*}{GFR}
&J-UNIWARD  & 45.38\%  & 37.63\% & 29.56\% & 21.65\% & 15.01\%\\
\cdashline{2-7}
&J-MSUNIWARD  & 45.52\%& 38.34\% & 30.65\%& 22.99\% & 16.20\%\\
&MSUERD\_SPA & 45.84\%& 39.64\% & 32.99\%& 26.26\% & 19.59\%\\\cdashline{2-7}
&\textbf{JEC-RL} & \textbf{46.75\%} & \textbf{41.20\%} & \textbf{35.85\%} & \textbf{29.10\%} & \textbf{22.77\%}\\

\midrule

\multirow{4}{*}{SCA-GFR}
&J-UNIWARD  & 42.39\%  & 33.61\% & 25.35\% & 18.29\% & 12.86\%\\
\cdashline{2-7}
&J-MSUNIWARD  & 42.20\%& 33.69\% & 25.73\%& 18.70\% & 13.23\%\\
&MSUERD\_SPA & 39.54\%& 31.49\% & 23.92\%& 18.27\% & 13.71\%\\\cdashline{2-7}
&\textbf{JEC-RL} & \textbf{44.99\%} & \textbf{39.16\%} & \textbf{32.94\%} & \textbf{26.65\%} & \textbf{21.13\%}\\

\midrule

\multirow{4}{*}{J-XuNet}
&J-UNIWARD  & 40.82\%  & 28.03\% & 19.67\% & 12.37\% & 7.59\%\\
\cdashline{2-7}
&J-MSUNIWARD  & 40.68\%& 28.64\% & 20.11\%& 13.01\% & 8.09\%\\
&MSUERD\_SPA & 29.80\%& 18.82\% & 11.93\%& 8.08\% & 4.46\%\\\cdashline{2-7}
&\textbf{JEC-RL} & \textbf{45.94\%} & \textbf{33.93\%} & \textbf{24.70\%} & \textbf{17.22\%} & \textbf{12.78\%}\\

\midrule

\multirow{4}{*}{SRNet}
&J-UNIWARD  & 31.86\%  & 18.68\% & 10.74\% & 6.90\% & 3.71\%\\
\cdashline{2-7}
&J-MSUNIWARD  & 32.10\%& 18.40\% & 11.21\%& 6.98\% & 3.91\%\\
&MSUERD\_SPA & 22.53\%& 11.62\% & 6.54\%& 3.93\% & 2.24\%\\\cdashline{2-7}
&\textbf{JEC-RL} & \textbf{39.36\%} & \textbf{24.29\%} & \textbf{16.08\%} & \textbf{9.89\%} & \textbf{6.55\%}\\
\bottomrule
\end{tabular}
\vspace{-0.6cm}
\end{table*}

\subsection{Settings}
\label{sec:setting}

\subsubsection{Steganographic methods}
\label{sec:steganographic methods}
Four steganographic methods {that can generate embedding costs from scratch}
were tested in our experiments, including J-UNIWARD (JPEG-Universal Wavelet Relative Distortion)\cite{UNIWARD}, J-MSUNIWARD (Microscope JPEG-Universal Wavelet Relative Distortion)\cite{MS}, MSUERD\_SPA (Microscope Uniform Embedding Distortion Revisited Filtering in Spatial Domain)\cite{MS}, and our proposed JEC-RL.
Being an effective traditional method, J-UNIWARD can be regarded as a kind of baseline, and J-MSUNIWARD is an improved version of J-UNIWARD, while MSUERD\_SPA is an improved version of UERD.
{Those methods that depend on pre-defined embedding costs, such as non-additive distortion methods and AE-based methods, were not included for comparison.
Since JS-GAN \cite{JSGAN} does not outperform J-UNIWARD, we did not include it neither.}
The payload is measured by bits per non-zero AC coefficient (bpnzAC) as in \cite{UNIWARD,UERD,MS}.
The optimal embedding simulator \cite{simulator} was employed for generating stego images.

The settings of JEC-RL are as follows.
The batch size $N$ was set as 24. The weighted parameters in \eqref{eq:GenLoss} were set as $\alpha=1$ and $\beta=10^{-7}$, respectively. The reward magnitude was set as $\xi=10^{7}$.
For all batch normalization layers, the momentum for the moving average was set as 0.999.
The Adam optimizer was used for optimization, wherein the learning rate was initialized as 0.0001 and it was decayed to $10\%$ in every 30,000 iterations.
The models at the 90,000-th training iteration under different payload rates were respectively used to calculate the embedding costs.

\subsubsection{Steganalyzers}
\label{sec:steganalyzer}
Five different steganalyzers were used to evaluate the security performance of the steganographic methods, including two feature-based methods, i.e., GFR (Gabor Filters Residual)\cite{GFR} and PHARM (PHase-Aware pRojection Model) \cite{PHARM}, one feature-based method utilizing selection channel information, i.e., SCA-GFR (Selection Channel Aware-Gabor Filters Residual)\cite{SCA}, and two CNN-based methods, i.e., J-XuNet \cite{XuJPEG} and SRNet \cite{SRNet}.
Security performance is evaluated as the detection error rate $P_{\text E}$ on the testing set, which is the average of the false alarm rate $P_{\text{FA}}$ and the missed detection rate $P_{\text{MD}}$:
\begin{equation}\label{eq:PE}
  P_{\text E} = \min \limits_{P_{\text{FA}}} \frac{1}{2}(P_{\text{FA}} + P_{\text{MD}}(C)).
\end{equation}

\subsubsection{Image Set}
Three image sets were used in experiments.

\begin{itemize}
   \item $\textit{Basic500k}$ ($256\times256$): it consists of 500,000 images, which were obtained by randomly selecting images with size larger than 256$\times$256 from ImageNet \cite{ImageNet} and then cropping their top-left 256$\times$256 regions. The images were further converted to grayscale and compressed with JPEG  quality factor 75. This image set has been used in \cite{zeng} and \cite{ADVEMB}.

   \item $\textit{BOSSBase}$ ($256\times256$): it consists of 10,000 images, taken from $\textit{BOSSBase}$ v1.01 image set \cite{BOSS} and down-sampled from $512\times512$ to $256 \times 256$ using the ``imresize'' Matlab function with \textit{Bicubic} Kernel. Then, the images were compressed into JPEG format.
   \item $\textit{BOWS2}$ ($256\times256$): it consists of 10,000 images. The original $512\times512$ images \cite{BOW} were  resized to $256\times256$ via the ``imresize'' Matlab function with \textit{Bicubic} Kernel. Then, the images were compressed into JPEG format.
\end{itemize}

In the experiments, the $\textit{Basic500k}$ was always used to train the JEC-RL.
$\textit{BOSSBase}$ and $\textit{BOWS2}$ were used to train the steganalyzers and evaluate the security steganographic performance.
Considering that different methods may require different amount of data for training, we treat feature-based and CNN-based methods differently.
Specifically, for feature-based methods such as GFR \cite{GFR}, PHARM \cite{PHARM}, and SCA-GFR \cite{SCA}, the images from $\textit{BOSSBase}$ were randomly split into a training set and a testing set with a proportion of $1:1$.
For CNN-based methods such as J-XuNet \cite{XuJPEG} and SRNet \cite{SRNet}, the images from $\textit{BOSSBase}$ were randomly split into a training set, validation set and a testing set with a proportion of $4:1:5$, and all images from $\textit{BOWS2}$ were included into the training set.
Note that the testing set was the same for  feature-based steganalyzers and CNN-based steganalyzers.

\subsection{Security Performance Against Steganalysis}
\label{sec:performance}
In this part, we compare the security performance of different steganographic methods.
The experimental results under state-of-the-art steganalyzers
are shown in Table \ref{tab:basic500k}.
The following observations can be made.

\begin{itemize}
  \item Compared with the baseline J-UNIWARD, JEC-RL has greatly improved the security performance. For example, in the case of 0.5 bpnzAC, the improvement is $8.58\%$, $7.76\%$, $8.27\%$, $5.19\%$, and $2.84\%$ against PHARM, GFR, SCA-GFR, Xu-Net, and SRNet, respectively.

  \item Facing with the feature-based steganalyzers without selection channel information, MSUERD\_SPA performs the best among the three conventional steganographic methods. In this case, the improvement over MSUERD\_SPA is $0.91\%$, $1.56\%$, $2.86\%$, $2.84\%$, and $3.18\%$ against GFR at 0.1 to 0.5 bpnzAC, respectively.

  \item When countering against the SCA-GFR and the CNN-based steganalyzers, J-MSUNIWARD is the best-performed conventional method. In such detection scenarios, JEC-RL can also achieve better security performance. For example, at 0.5 bpnzAC, the improvement is $7.90\%$, $4.69\%$, and $2.64\%$ against SCA-GFR, J-XuNet, and SRNet, respectively.

  \item Even in a low payload situation such as 0.1 bpnzAC, JEC-RL can obtain significant improvement. When comparing with J-UNIWARD, the improvement against GFR and PHARM is around $1\%$. The improvement against more advanced CNN-based steganalyzers, i.e., J-XuNet and SRNet, is even more significant, which is more than $5\%$ and $7\%$, respectively.

\end{itemize}

It can be concluded that although the best conventional method may be different against different steganalyzers,
our proposed JEC-RL can always outperform the best one.

\begin{figure}[t!]
\center
\includegraphics[width=0.44\textwidth]{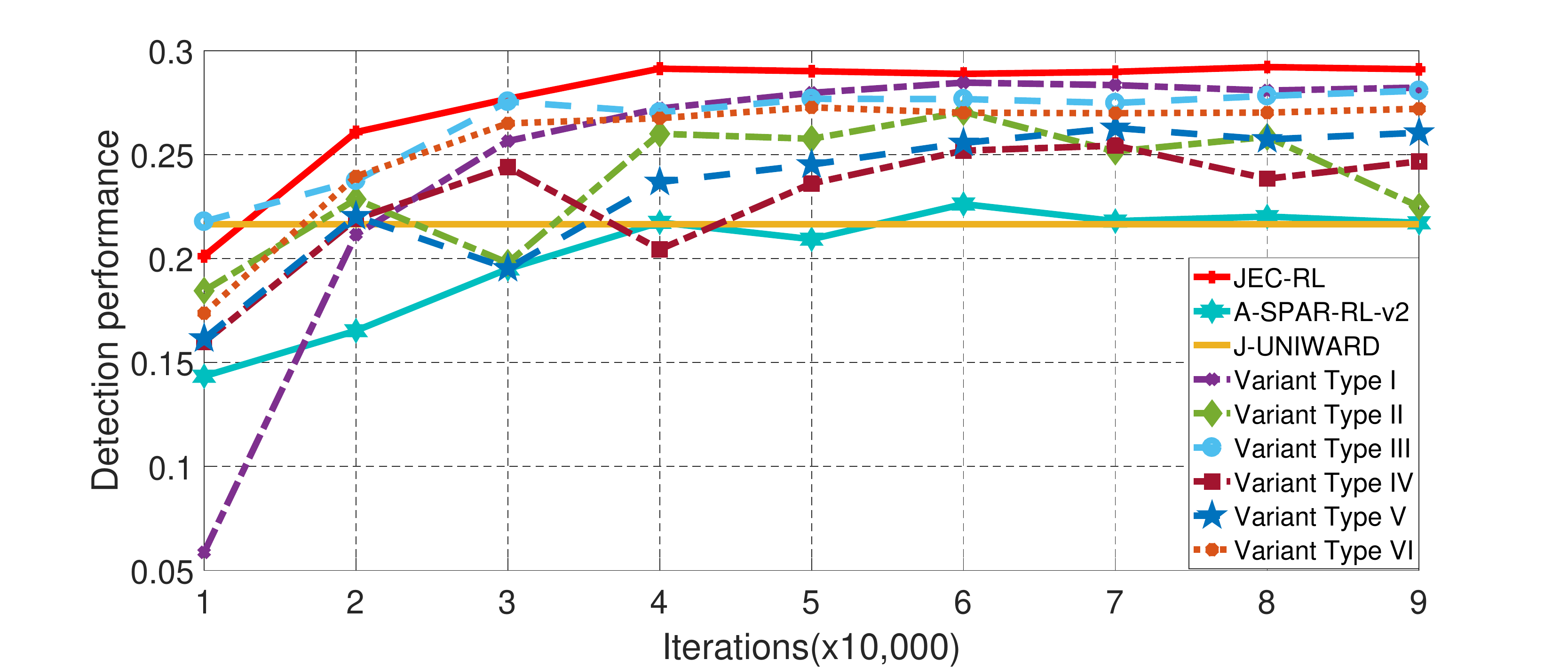}
\label{fig:stable1}
\caption{Security performance of JEC-RL and its variants against GFR steganalyzer on 0.4 bpnzAC.}
\label{fig:ablation}
\vspace{-0.6cm}
\end{figure}

\subsection{Ablation Studies on Network Architecture}
\label{sec:ablation}

In this part, we conduct ablation studies on the proposed network architecture. Specifically, we take JEC-RL as a baseline, and investigate the impact of the first module of the policy network by Variant Type I, different kinds of preprocessing layer in the environment network by Variant Type II, III, and IV, and the impact of the depth and width of environment network by Variant Type V and VI.

\subsubsection{Variant Type I (learning block-level texture complexity
in the first module of policy network)}
The first module of the policy network applied in JEC-RL evaluates the pixel-level texture complexity.
In Variant Type I, the first module is
used to learn block-level texture complexity instead of pixel-level texture complexity.
To this end, for an input image with size $H \times W$,
the last layer of the first module outputs $H/8\times W/8 \times 64$ feature maps, rather than the $H\times W \times 1$ feature map in the original JEC-RL.
The second module takes such $H/8\times W/8 \times 64$ feature maps as input, and utilizes three convolutional groups to obtain $\mathbf{F^{\prime\prime\prime}} = ({f}^{\prime\prime\prime}_{i,j,d})^{H/8 \times W/8 \times 64}$, wherein all convolutional kernels are of size $3\times3\times64\times64$ and all convolutional strides equal to 1.

\subsubsection{Variant Type II (utilizing $4\times4$ DCT basis filters in the preprocessing layer of environment network)}
In Variant Type II, the $8\times8$ DCT basis filters were replaced with $4\times4$ DCT basis filters, which were applied in J-XuNet.

\subsubsection{Variant Type III (utilizing SRM filters in the preprocessing layer of environment network)}
In Variant Type III, the $8\times8$ DCT basis filters were replaced with 30 high-pass filters from SRM \cite{SRM}, which were adopted in SPAR-RL-v2.

\subsubsection{Variant Type IV (utilizing learnable filters in the preprocessing layer of environment network)}
In Variant Type IV, the $8\times8$ DCT basis filters were replaced with learnable convolutional filters, which were adopted in SRNet.

\subsubsection{Variant Type V (increasing the depth of environment network)}
Considering that deeper network architecture is preferable for better steganalytic detection performance.
In Variant Type V, the learnable part of the environment network (6-layer) was replaced with the 22-layer in J-XuNet.

\subsubsection{Variant Type VI (decreasing the width of environment network)}
In JEC-RL, the learnable part of the environment network is a width-expanded version of that in XuNet. In Variant Type VI, the original setting of network width in XuNet was adopted.

The experimental results on 0.4 bpnzAC with a GFR
steganalyzer are shown in Fig. \ref{fig:ablation}.
It can be observed that the JEC-RL outperforms conventional method, previous cost learning method, and its variants.
The following conclusions can be made.
\begin{itemize}
  \item {It can be observed that the proposed JEC-RL and all of its variants outperform J-UNIWARD and A-SPAR-RL-v2, verifying the effectiveness of the domain-transition paradigm based policy network and the gradient-oriented environment network.}

  \item To learn more effective embedding policies, learning fine-grained pixel-level texture complexity features is more effective than learning block-level features in the first part of the policy network.

  \item The $8\times8$ DCT basis filters may be the most suitable one in the environment network's preprocessing layer.
      The reason is that the $8\times8$ DCT basis filters
      filters have a preference in propagating gradients to modifications on the DCT modes that are close to their own frequencies.
      By contrast, the $4\times4$ DCT basis filters are coarse-grained and cannot cover the whole frequency band for 64 DCT frequency-modes, and the spatial SRM filters or learnable filters cannot perfectly match 64 JPEG DCT frequency-modes.
      More detailed analysis of the $8\times8$ DCT basis filters is given in Section \ref{sec:investigation}.

  \item To improve the environment network's capability, increasing the network width is more effective than increasing the network depth. Although a deeper CNN can usually obtain better security performance, it may lead to potential issues in inefficient gradients.
      Therefore, increasing the depth of the environment network immoderately may degrade its performance in reward assignment for obtaining optimal embedding policies.

\end{itemize}

\subsection{Incorporating Texture Calculation Process in Traditional Steganographic Methods into Texture Evaluation Module}
\label{sec:refine}
{In this part, we try to incorporate the texture calculation process into the first module of the policy network, and then continue to utilize the two rest modules for cost learning.
We take J-UNIWARD and MSUERD\_SPA for example, and abbreviate the scheme as
JEC-RL(J-UNI) and JEC-RL(MSU).}

%
In JEC-RL(J-UNI), the first module directly outputs the pixel-level texture complexity matrix $\mathbf T = (t_{i,j})^{H \times W\times3 }$, wherein each channel is obtained by convolving the decompressed JPEG image with one of the three Daubechies wavelet filters, as it did in J-UNIWARD.
In JEC-RL(MSU), the first module firstly obtains
block-level texture $\mathbf T^{\prime} = (t_{a,b}^{\prime})^{H/8 \times W/8 }$ as the weighted sum of the block energy $E_{a,b}$ wherein $t_{a,b}^{\prime} = E_{a,b}+0.25\cdot \sum_{\hat{E}\in \mathbb{\hat{E}}_{a,b}}\hat{E}$, as it did in both MSUERD\_SPA and UERD shown in \eqref{eq:E}. Then, the block level texture matrix is  nearest-neighbor upsampled to pixel-level texture $\mathbf T = (t_{i,j})^{H \times W}$ by a three layer learnable neural network.
{Since their texture calculation process contains none or few learnable parameters compared with a pixel-to-pixel CNN structure, a deeper network with six convolutional groups is used in the second module to facilitate effective learning, where the odd groups have a convolution stride of $1\times1$ and the even groups have a convolution stride of $2\times2$}.}
The experimental results are shown in Table \ref{tab:SPARRLJR}.
The following observations can be made.

\begin{enumerate}
  \item JEC-RL(J-UNI) outperforms J-UNIWARD under all circumstances. For example, the improvement on 0.3 bpnzAC is $4.33\%$, $7.96\%$, and $4.39\%$ against GFR, SCA-GFR and J-XuNet, respectively.

  \item MSUERD\_SPA outperforms J-UNIWARD against GFR, but is comparable or inferior to J-UNIWARD against SCA-GFR and J-XuNet.
      JEC-RL(MSU) can make up for the deficiency and outperforms MSUERD\_SPA against SCA-GFR and J-XuNet by $5.89\%$ and $3.69\%$.

  \item 
      JEC-RL(J-UNI) outperforms JEC-RL(MSU).
      The reason is that the pixel-level texture can be directly obtained in J-UNIWARD, while that is upsampled from block-level texture in MSUERD\_SPA, which may leave some important fine-grained information.
\end{enumerate}
It can be concluded that JEC-RL(J-UNI)/JEC-RL(MSU) can make better use of the same texture information, implying that the DCT feature extraction module can effectively capture inter-block and intra-block characteristics. The performance of the vanilla JEC-RL is superior to JEC-RL(J-UNI)/JEC-RL(MSU), indicating that the design paradigm can maximize its learning ability via end-to-end optimizing its modules.

\begin{table}[t!]
\renewcommand\arraystretch{1.1}
{\caption{
$P_{\text E}$ of steganographic methods with JPEG quality factor 75. The numbers in parentheses show the performance difference between JEC-RL(J-UNI)/JEC-RL(MSU) and J-UNIWARD/MSUERD\_SPA.}
\label{tab:SPARRLJR}}
\scriptsize
\centering
\begin{tabular}{ccll}
\toprule
\textbf{Stegan-} &
\textbf{Steganographic}&
\multirow{2}{*}{\textbf{0.3 bpnzAC}} & \multirow{2}{*}{\textbf{0.5 bpnzAC}}\\\textbf{alyzer}& \textbf{method}& &
\\\midrule

\multirow{5}{*}{GFR}
&J-UNIWARD  & 29.56\%  & 15.01\% \\
&\textbf{JEC-RL(J-UNI)} & 33.89\%(\textbf{$\uparrow$4.33\%})  & 19.35\%(\textbf{$\uparrow$4.34\%})\\\cdashline{2-4}
\cdashline{2-4}
&MSUERD\_SPA & 32.99\%& 19.59\% \\
&\textbf{JEC-RL(MSU)} &31.49\%({$\downarrow$1.50\%}) & 17.09\%({$\downarrow$2.50\%})\\\cdashline{2-4}
&JEC-RL  & 35.85\%  & 22.77\% \\\cline{1-4}

&J-UNIWARD  & 25.35\%  & 12.86\% \\
SCA-&\textbf{JEC-RL(J-UNI)} & 33.31\%(\textbf{$\uparrow$7.96\%})  & 20.25 \%(\textbf{$\uparrow$7.39\%})\\\cdashline{2-4}
\cdashline{2-4}
GFR&MSUERD\_SPA & 23.92\%& 13.72\% \\
&\textbf{JEC-RL(MSU)} &29.81\%(\textbf{$\uparrow$5.89\%}) & 16.17 \%(\textbf{$\uparrow$2.45\%})\\\cdashline{2-4}
&JEC-RL  & 32.94\%  & 21.13\% \\\cline{1-4}

\multirow{4}{*}{J-XuNet}&J-UNIWARD  & 19.67\%  & 7.59\% \\
&\textbf{JEC-RL(J-UNI)} & 23.96\%(\textbf{$\uparrow$4.29\%})  & 10.49\%(\textbf{$\uparrow$2.90\%})\\
\cdashline{2-4}
&MSUERD\_SPA & 11.93\%& 4.46\% \\
&\textbf{JEC-RL(MSU)} & 15.62\%(\textbf{$\uparrow$3.69\%}) & 6.21\%(\textbf{$\uparrow$1.75\%})\\
\cdashline{2-4}
&JEC-RL  & 24.70\%  & 12.78\% \\
\bottomrule
\end{tabular}
\end{table}

\begin{figure*}[t!]
\centering
\subfigure[$8\times8$ DCT basis filters.]
{\includegraphics[height=3.2cm]{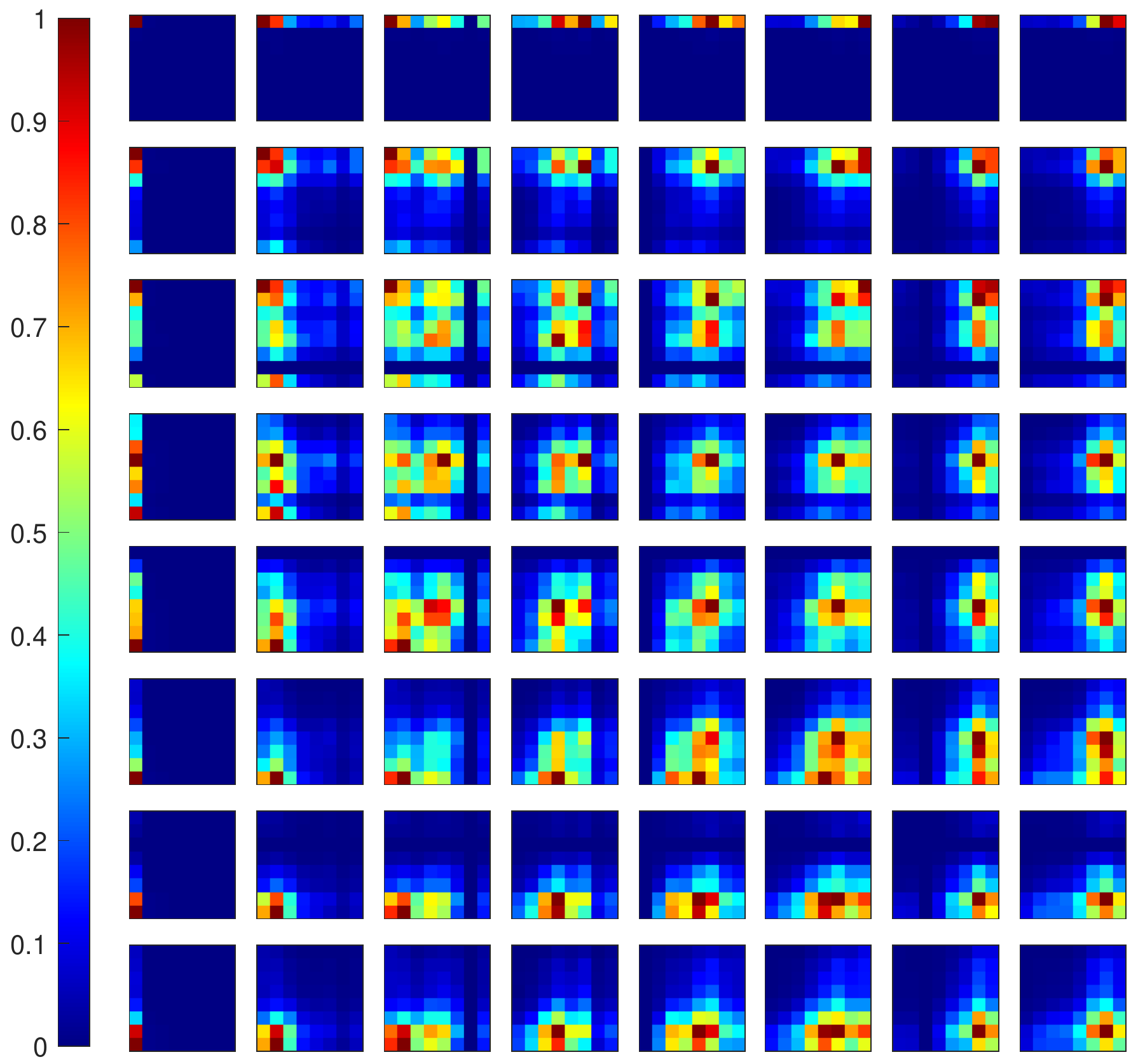}
\label{fig:64DCT_g}}
\hspace{.1in}
\subfigure[$4\times4$ DCT basis filters.]
{\includegraphics[height=3.2cm]{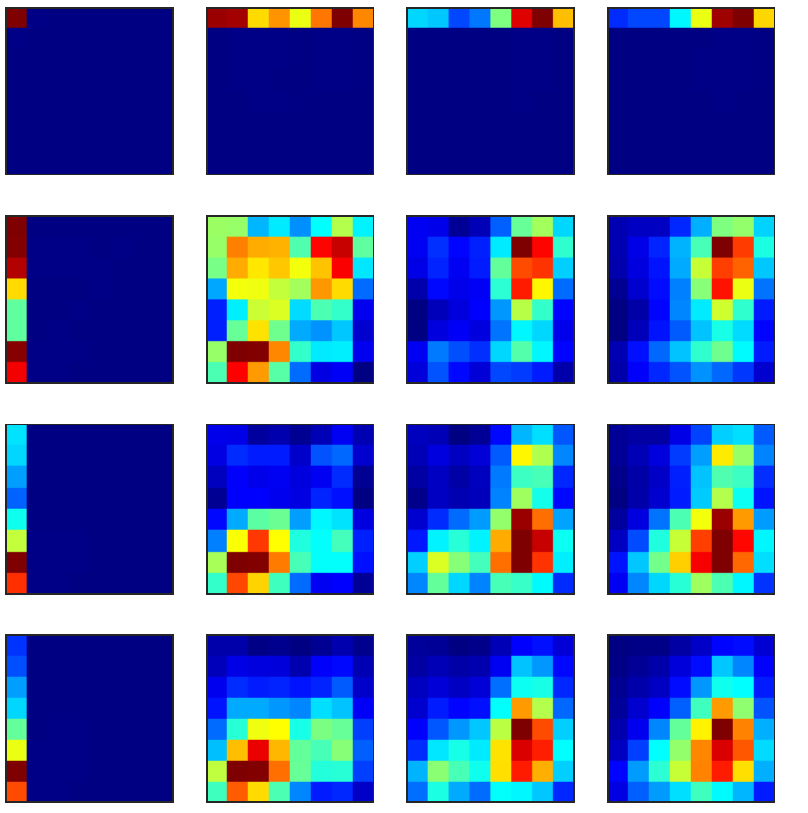}
\label{fig:16DCT_g}}
\hspace{.1in}
\subfigure[SRM high-pass filters.]
{\includegraphics[height=3.2cm]{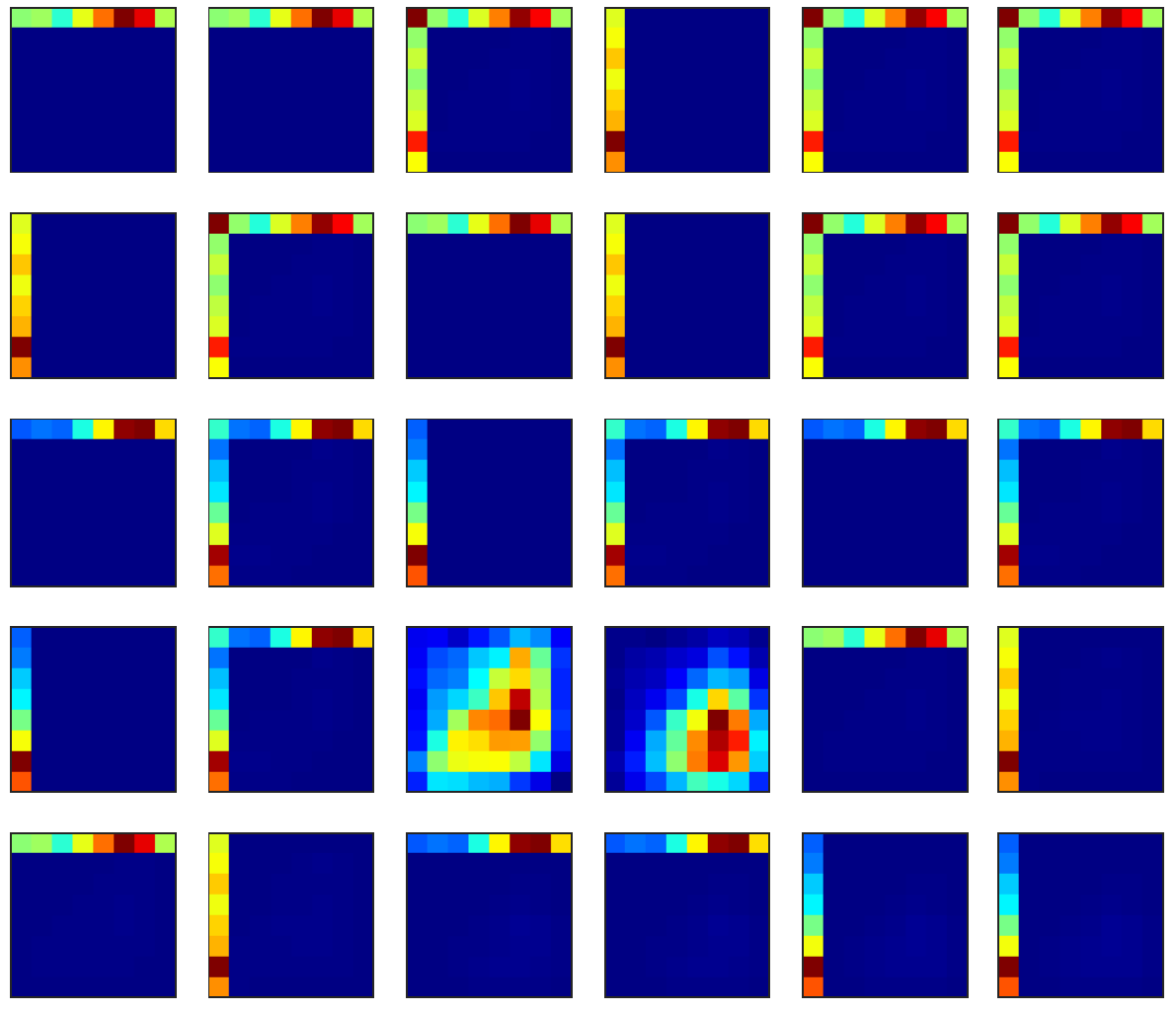}
\label{fig:30SRM_g}}
\hspace{.1in}
\subfigure[top-$n$-statistics.]{\includegraphics[height=3.2cm]{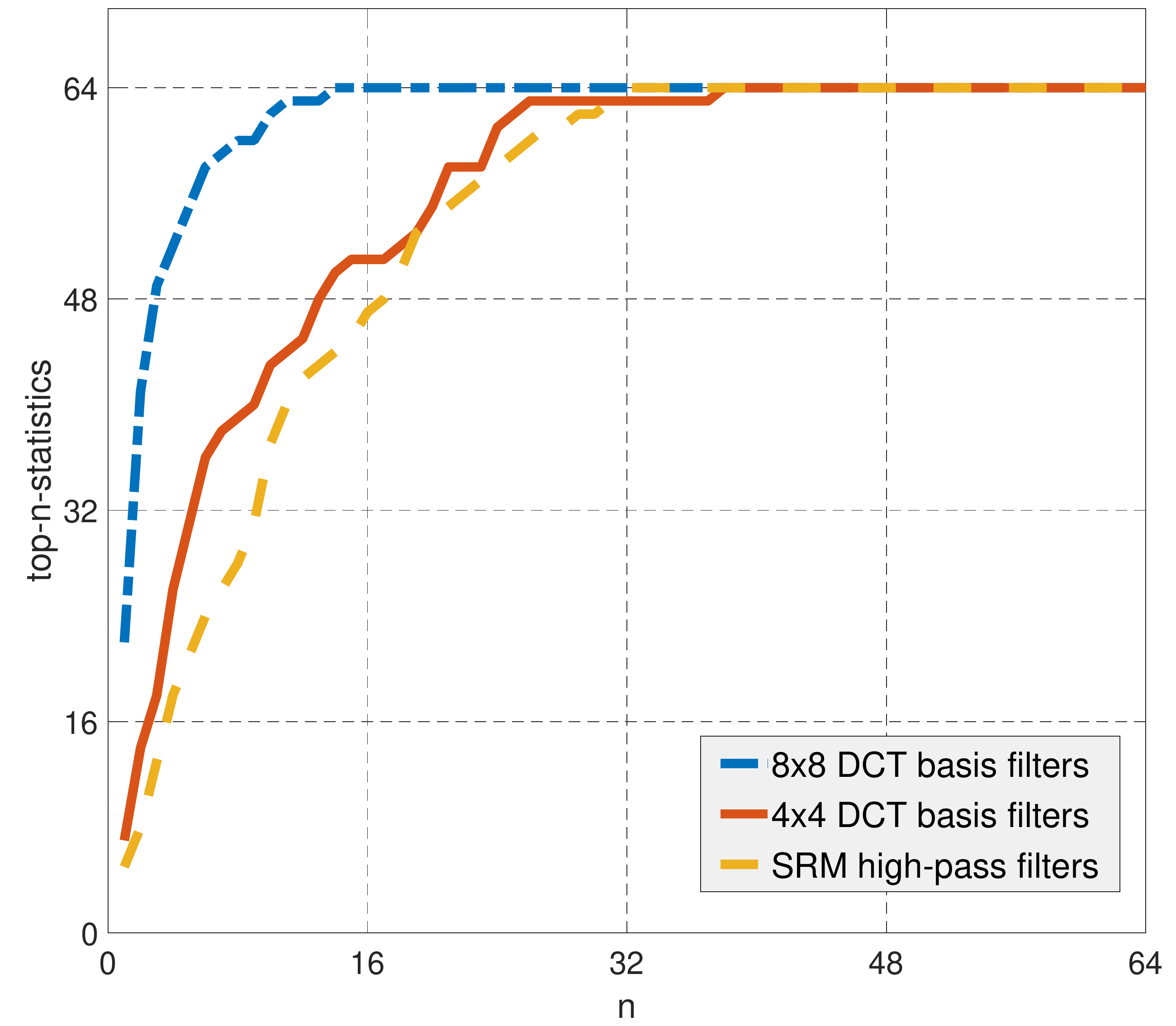}
\label{fig:select}}
\caption{The averaged accumulated gradient
component matrices ${\mathbf{E}}^{u,v} = ({e}^{u,v}(k,l))^{8 \times 8}$ $(1 \le k,l \le 8)$ over 1,000 images from $\textit{BOSSBase}$ for three filter sets ($8\times8$ DCT basis filters (a), $4\times4$ DCT basis filters(b), and 30 SRM high-pass filters(c)) and the curves of top-$n$-statistics $s_n$ (d). The elements within the same matrix are normalized to [0,1],
and the red hot color denotes large value while the blue cold color denotes small value in (a), (b) and (c).
}\label{fig:8x8}
\end{figure*}

\subsection{Investigation on the Role of Filters for Gradient Propagation}
\label{sec:investigation}


In this part, we further investigate the role of the filters in the preprocessing layer of JEC-RL
for propagating gradients in reward assignment.
%
In the backpropagation process, the gradients of the environment network's loss function are propagated through the filters in the preprocessing layer to the modifications.
We intend to investigate
how the gradients are propagated towards each DCT mode through each filter in a
filter set. We compare three filter sets, i.e., $8\times8$ DCT basis filters, $4\times4$ DCT basis filters, and 30 SRM high-pass filters.



Denote a filter $\mathbf Z^{u,v}$, where $(u,v)$
 is the index in a filter set and specifically
 $1 \le u,v\le 8$ for $8\times8$ DCT basis filters,
 $1 \le u,v\le 4$ for $4\times4$ DCT basis filters, and
 $1 \le u\le 5, 1 \le v\le 6$ for  SRM filters, respectively.
 $(u,v)$ also represents the corresponding frequency for DCT basis filters.
%
The feature map obtained by processing an input image  $\mathbf{X}^{J}$ with the filter $\mathbf Z^{u,v}$ is denoted as $\mathbf S^{u,v} = (s_{i,j}^{u,v})^{H \times W}$.
The modification map is denoted as $\mathbf{M}^{J} = (m_{a,b}^{k,l})^{H \times W }$, where $(k,l)$ is the position of the DCT frequency-mode and $(a,b)$ is the position of the DCT block.
The gradient which we investigate is $\frac{\partial{s_{i,j}^{u,v}}}{\partial{m_{a,b}^{k,l}}}$
for that it can be regarded as the feedback route from the image residuals to the modifications.
We form an \textit{accumulated gradient component matrix} ${\mathbf{E}}^{u,v} = ({e}^{u,v}_{k,l})^{8 \times 8}$ ($1 \le k,l \le 8$))
for the $(u,v)$-th filter.
The $(k,l)$-th element in the matrix is obtained by
\begin{equation}
\label{eq:gradient}
  \begin{aligned}
    e^{u,v}_{k,l} = \sum_{a=1}^{H/8}\sum_{b=1}^{W/8}  \Big( \sum_{i=1}^{H}\sum_{j=1}^{W} \left| \frac{\partial{s_{i,j}^{u,v}}}{\partial{m_{a,b}^{k,l}}} \right | \Big).
  \end{aligned}
\end{equation}
In \eqref{eq:gradient}, the magnitude of the gradient components are first summed up
over all image residuals, and then summed up over all DCT blocks for the $(k,l)$-th DCT mode.
It may evaluate the overall effects of the gradients propagated
from all residuals towards the $(k,l)$-th DCT mode by the $(u,v)$-th filter.
The larger the element it is in the matrix,
the larger the accumulated gradient components propagate back to the corresponding
DCT mode through such a filter.

{We calculated the averaged value of ${\mathbf{E}}^{u,v}$ over 1,000 images from $\textit{BOSSBase}$ and
normalized the elements in each ${\mathbf{E}}^{u,v}$ to [0,1].
The results for the three filter sets can be visualized in Fig. \ref{fig:8x8}(a) to \ref{fig:8x8}(c).
It can be observed that the filters in a given filter set
may have a preference in propagating gradients over different DCT modes.
It is interesting to see that
 $8\times8$ DCT basis filters and $4\times4$ DCT basis filters
have a preference in serving the DCT modes that are close to their own frequencies,
while most of SRM filters show a kind of preference in serving the DCT modes in the horizontal or vertical direction.
To further analyze how well each DCT mode can receive large accumulated gradients, we performed the following statistical analysis.
}


\begin{enumerate}
  \item For each ${\mathbf{E}}^{u,v} = ({e}^{u,v}_{k,l})^{8 \times 8}$, sort its 64 elements ${e}^{u,v}_{k,l}$ in a descending order, and the corresponding sorting order of ${e}^{u,v}_{k,l}$ is denoted as ${o}^{u,v}_{k,l} \in \{1, 2, \cdots, 64\}$.
      The larger the element, the smaller the order.
  \item
  In a given filter set with a number of $U \times V$ filters\footnote{We have $U=V=8$ for $8 \times 8$ DCT basis filters, $U=V=4$ for $4 \times 4$ DCT basis filters, and $U=5$, $V=6$ for SRM filters.)}, count how many times a DCT mode $(k,l)$ $(1 \le k,l \le 8)$ ranks among the top-$n$ in the sorting order, called
  top-$n$-rate, as
      \begin{equation}
      \begin{aligned}
        r_{k,l,n} = \sum_{u=1}^{U}\sum_{v=1}^{V} \delta({o}^{u,v}_{k,l} \leq n) \quad  (1 \le k,l \le 8).
      \end{aligned}
       \end{equation}
\item Then, for a given $n$,
  count the amount of DCT modes that can attain non-zero top-$n$-rate, called top-$n$-statistics, as
      \begin{equation}
      \begin{aligned}
         s_n = \sum_{k=1}^{8}\sum_{l=1}^{8}\delta(r_{k,l,n}>0).
      \end{aligned}
      \end{equation}
\end{enumerate}
Fig. \ref{fig:select} shows the curve of $s_n$ ($0 \le n \le 64$) for the three filter sets.
It can be observed that
for a given $n$, the $8\times8$ DCT basis filter set can always obtain a top-$n$-statistics no less than the other two filter sets,
{indicating that the filters within such filter set are more likely to be responsible for propagating gradients to each of the 64 DCT modes.}

\begin{figure}[t!]
\centering
\subfigure[Cover image]{\includegraphics[width=2.4cm]{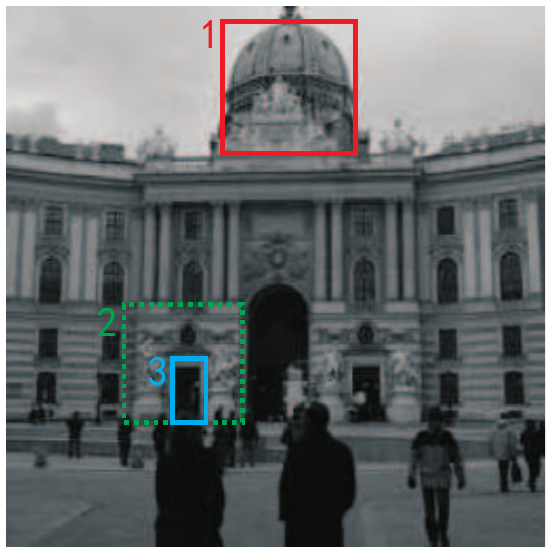}
\label{fig:cover}}
\subfigure[MP of J-MSUNIWARD]
{\includegraphics[width=2.4cm]{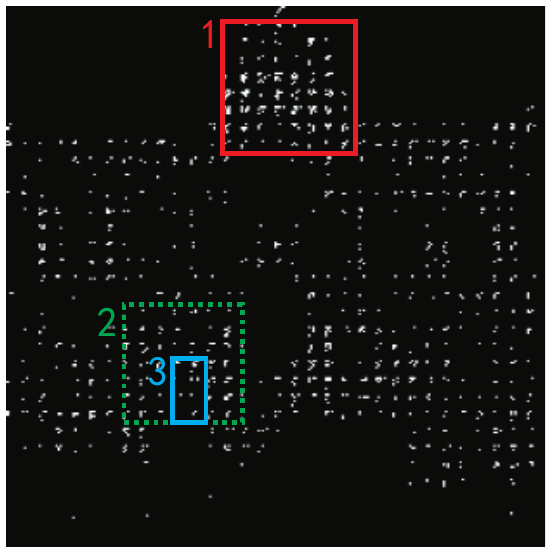}
\label{fig:m_JMSUNIWARD}
}
\subfigure[MP of MSUERD\_SPA]
{\includegraphics[width=2.4cm]{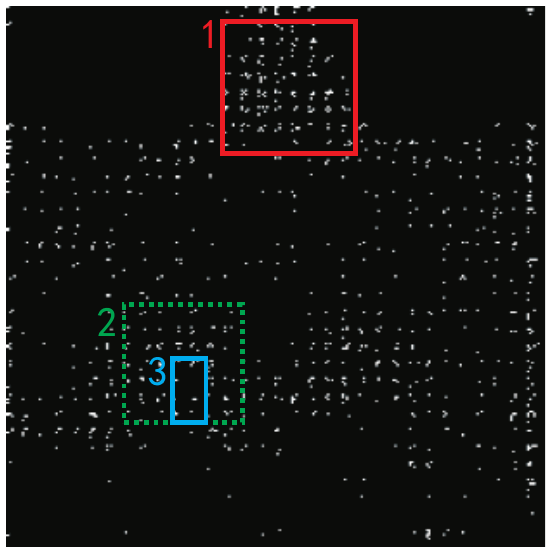}
\label{fig:m_MSUERDSPA}
}
\hspace{.01in}
\subfigure[MP of A-SPAR-RL-v2]
{\includegraphics[width=2.4cm]{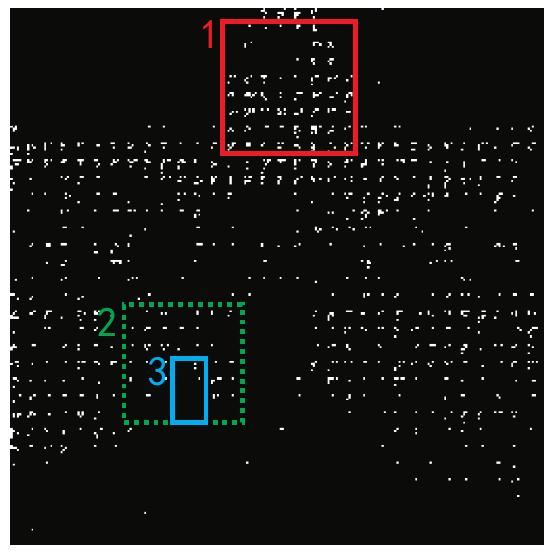}
\label{fig:m_ASPARRLv2}
}
\hspace{.01in}
\subfigure[MP of JEC-RL]{\includegraphics[width=2.4cm]{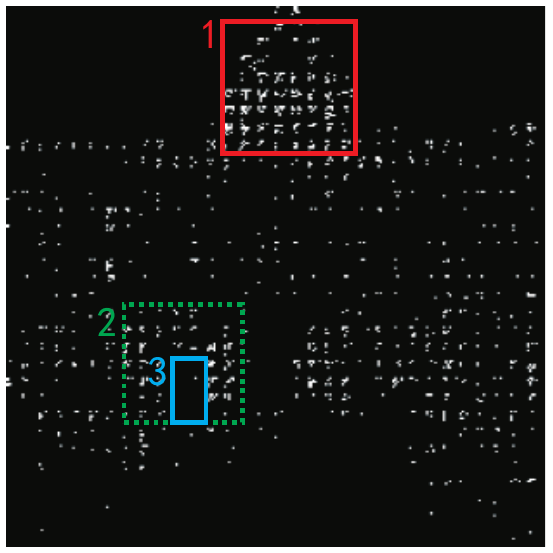}
\label{fig:m_SPARRLJ}
}
\subfigure[ECP of JEC-RL]{\includegraphics[width=2.4cm]{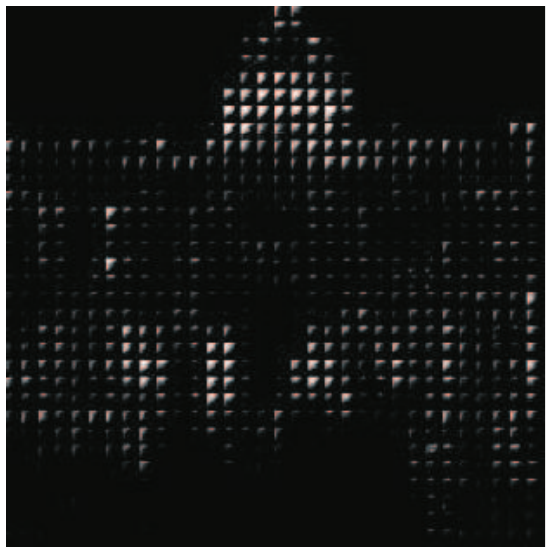}
\label{fig:prob}
}
 {\caption{Illustration of the cover image ``01013.jpg" from $\textit{BOSSBase}$, the modification map (MP) of different steganographic methods, and the embedding change probabilities (ECP) of JEC-RL on 0.4 bpnzAC.}
 \label{fig:modification}}
\end{figure}

\subsection{Analyzing the Embedding Patterns}
\label{sec:visulazation}

The  modification map and the embedding change probabilities of a cover image are visualized in Fig. \ref{fig:modification}.
To better illustrate the modification map, we select some regions from a sample image.
Box 1 contains the regions with high texture complexity.
As for Box 2, those outside Box 3 are texture regions, while those inside Box 3 are smooth regions.
It can be observed that compared with the state-of-the-art traditional methods J-MSUNIWARD and MSUERD\_SPA, and A-SPAR-RL-v2 adapted from spatial domain,
the embedding pattern of JEC-RL is more likely to concentrate in texture regions such as those inside Box 1, and avoid to spread in the smooth regions such as those inside Box 3.

It can be observed that the embedding change probabilities obtained by JEC-RL have obvious
adaptivity on image content as well as on different DCT frequency-mode positions.
Specifically, the embedding change probabilities are mostly concentrated in the texture region of an image. Besides,
the embedding change probabilities
in the upper-left corner of DCT block is larger than those in the bottom-right corner.
It is widely acknowledged that in JPEG steganography the DCT coefficients in low or medium frequency-modes are more suitable to be modified than those in high frequency-modes.
As a result, most of the traditional methods are designed according to such heuristic rule.
This rule can also be verified  by JEC-RL,
where the embedding change probabilities are automatically learned via the interactions between the policy network and the environment network in a data-driven manner.

\section{Conclusion}
\label{sec:conclusion}
In this paper, we have proposed an automatic cost learning method for JPEG images called JEC-RL, wherein the policy network and environment network are specifically designed according to JPEG domain knowledge.
 Within the policy network, its first module outputs the pixel-level texture complexity, and then its second module generates DCT features, and finally the third module rearranges the DCT features into embedding policies represented in a $8\times8$ DCT frequency-mode structure.
As for the environment network, the preprocessing layer, the network depth, and the network width are carefully designed for the purpose of efficient gradient propagation for reward assignment.
{We stress that DCT basis filters are more suitable than high-pass filters in the preprocessing layer to provide sufficient frequency resolution
due to fact that embedding modifications are possible to be taken place in all DCT modes.}
Extensive experiments have demonstrated that JEC-RL can automatically learn outstanding JPEG embedding costs, and the following conclusions can be made.

\begin{enumerate}
  \item Compared to the traditional steganographic methods, JEC-RL has achieved the state-of-the-art security performance against different advanced steganalyzers.
  \item The network architecture applied in JEC-RL
  {has been deliberately designed according to some JPEG DCT characteristics, including the texture level of DCT blocks, the correlation among DCT coefficients, and the different impacts of DCT frequency-modes. These considerations
  have significant positive impacts on security performance.} Ablation studies have shown the effectiveness of the proposed three-module composed policy network and the gradient-oriented environment network equipped with the $8\times8$ DCT basis filters.
  \item JEC-RL can be used to work with traditional methods such as J-UNIWARD and MSUERD\_SPA. Based on the same kind of texture complexity information, JEC-RL(J-UNI) and JEC-RL(MEU) can obtain more effective steganographic embedding costs than J-UNIWARD and MSUERD\_SPA.
\end{enumerate}

To the best of our knowledge, this paper is the first work that can learn JPEG embedding costs from scratch and outperform traditional cost functions.
{
To summarize, besides of using more advanced network structure for learning automatically,
the domain knowledge also plays an important role and should be exploited
in the task-specific schemes.
}

To further improve the performance, the following aspects may worth investigation. Firstly, it is interesting to learn embedding costs for non-additive distortion. For example, the policy network may generate joint embedding policies for DCT coefficients with high correlation, as it did in \cite{decompose} for joint distortion.
Secondly, the side information from uncompressed images \cite{sideinfor} may be taken into account in the policy network to obtain more secure modification actions.
{Thirdly, more attentions should be paid to the reward function.
For example, multiple environment networks can be involved to yield rewards in a way similar as min-max strategy \cite{minmax}. Besides, the rewards may also be designed in a way independent of the the gradients.}
Lastly, we hope to extend our cost learning method to other media carriers such as video or audio \cite{audio}.

\section{appendix}
\label{sec:appendix}

\begin{figure}[t]
\centering
\subfigure[Pixel-level texture complexity evaluation module.]{\includegraphics[height=6cm]{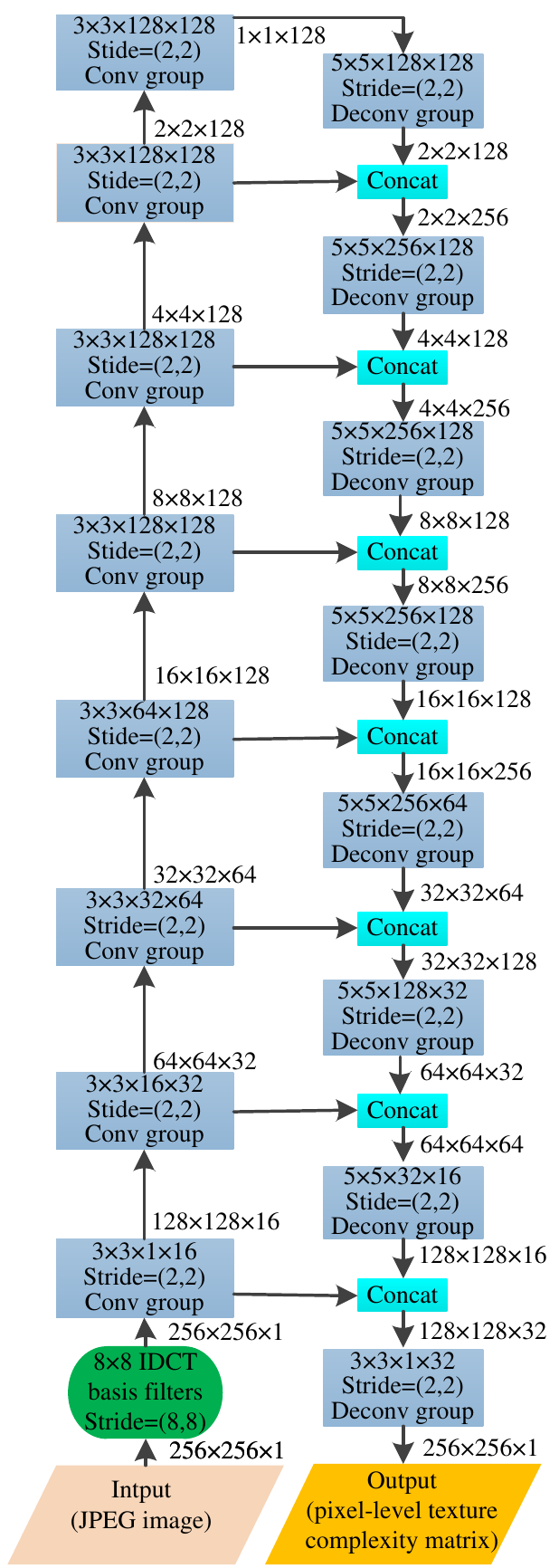}
\label{fig:fp}}
\hspace{.1in}
\subfigure[DCT feature extraction module.]{\includegraphics[height=7cm]{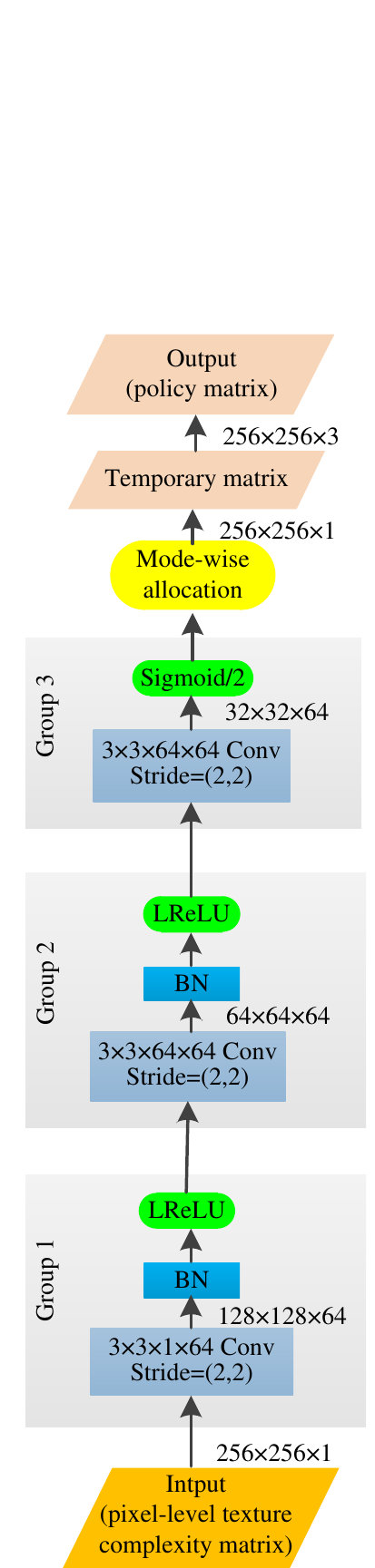}
\label{fig:sp}}
\hspace{.1in}
\subfigure[Gradient-oriented environment network.]{\includegraphics[height=6cm]{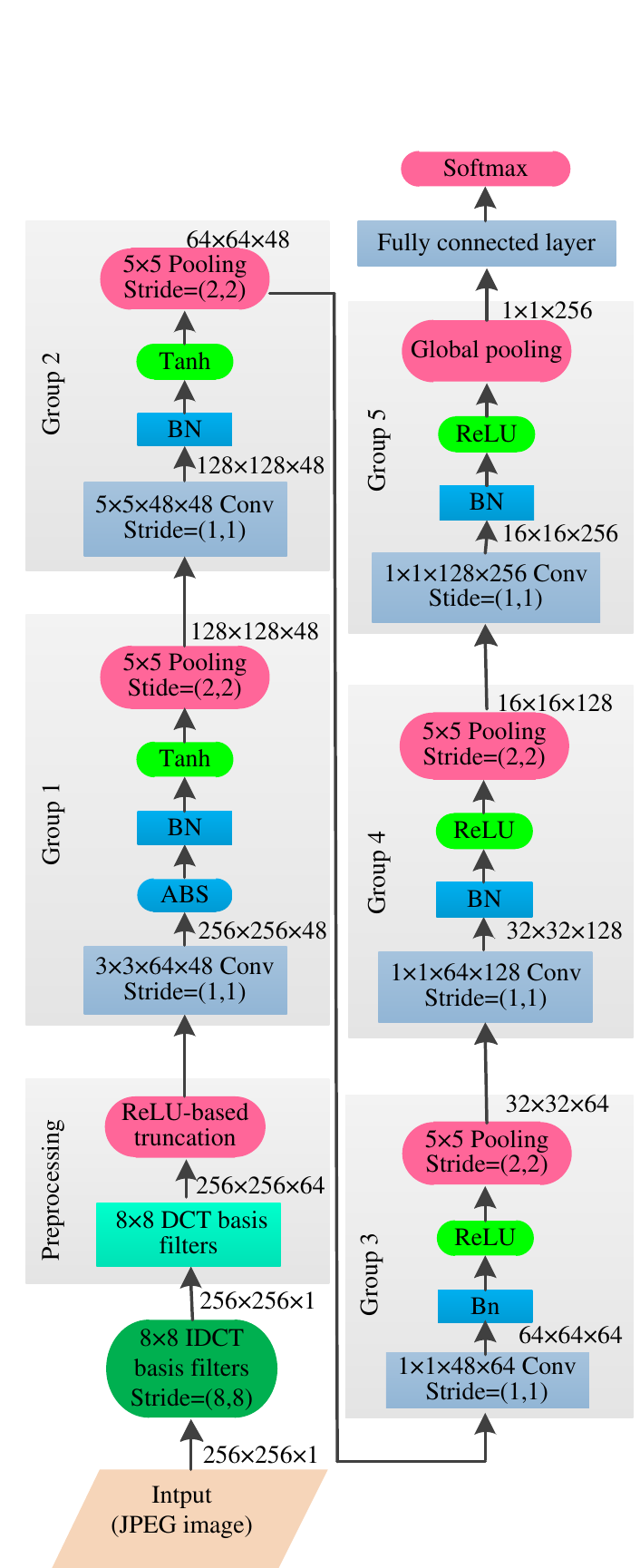}
\label{fig:e}}
\caption{
Network structures used in JEC-RL.
}\label{fig:Architecture}
\end{figure}

In this appendix, we show the architecture of the learnable part in JEC-RL in Fig. \ref{fig:Architecture}, including the pixel-level texture complexity evaluation module and the DCT feature extraction module in the policy network, and the gradient-oriented environment network.
The kernel configurations of the convolutional layers are given in the format: kernel width $\times$ kernel height $\times$ number of input feature maps $\times$ number of output feature maps. The sizes of the output
feature maps are given in the format: height $\times$ width $\times$ number of feature maps. In the first part of the policy network, each convolutional group consists of a convolutional layer, a batch normalization layer, and a ReLU activation function, while each deconvolutional group consists of a deconvolutional layer, a batch normalization layer, and a leaky ReLU activation function.

\normalem
\bibliographystyle{IEEEtran}
\bibliography{references, IEEEabrv}

\end{document}